# Generalized oscillator strength of endohedral molecules[1]

**M. Ya. Amusia**[1, 2], **L. V. Chernysheva**[2], **and E. Z. Liverts**[1]

[1]Racah Institute of Physics, the Hebrew University, Jerusalem 91904, Israel
[2]Ioffe Physical-Technical Institute, St.-Petersburg 194021, Russia

**Abstract**

We investigate here the fast electron scattering upon endohedral atoms that present a fullerene $C_N$ staffed by an atom A, $A@C_N$. We calculate the inelastic scattering cross-section expressing it via generalized oscillator strengths (GOS) density. We take into account two major effects of $C_N$ upon ionization of the atom A. Namely, the scattering of the electron by the static potential of the fullerenes shell and modification of the interaction between the fast incoming and atomic electrons due to polarization of the fullerenes shell by the incoming electron beam.

To obtain the main features of the effect, we substitute the complex fullerenes shell $C_N$ by a static zero thickness potential, express its deformation under incoming electron action via $C_N$ polarizabilities. We perform all consideration in the frame of the so-called random phase approximation with exchange (RPAE) that gave reliable results for GOSes of isolated atoms, and expressions for absolute and differential in angle cross-sections. We limit ourselves with dipole and biggest non-dipole contributions to the differential in angle cross-section.

We compare fast electron scattering with photoionization as a source of information on the target electronic structure and emphasize some advantages of fast electron scattering. As concrete objects of calculations, we choose noble gas endohedrals $Ar@C_{60}$ and $Xe@C_{60}$. The results are presented for two transferred momentum $q$ values: $q=0.1$ and $1.0$. Even for small $q=0.1$, in the so-called optical limit, the deviation from photoionization case is prominent and instructive.

As an interesting and very specific object, we study onion-type endohedrals $A@C_{N1}@C_{N2}$, in which the all construction $A@C_{N1}$ is stuffed inside $C_{N2}$ with $N_2 \gg N_1$.

## 1. Introduction

In this paper, we investigate fast electron[2] inelastic scattering upon endohedral atoms or endohedrals. This is a system that consist of a fullerenes shell $C_N$ and an atom A, stuffed inside. The notation for this object is A@$C_N$. Most simple for analysis is close to perfect spherically – symmetry fullerene $C_{60}$, which is able to cage almost any element of the periodic table and simple molecules. Noble gas atoms occupy their position at the center of $C_{60}$.

---

[1]This paper is dedicated to the eighties anniversary of Prof. I. Kaplan birthday. I (MYA) know him for about forty years. When we became acquainted, he was already a well-recognized expert in physical chemistry. We met each other very frequently at various conferences inside USSR. At that time, the way out of the "iron curtain" was locked for both of us. After collapse of the USSR, our traveling activity increased enormously, while meetings became, alas, much more seldom. Perhaps, it is inversely proportional to the increase of our traveling area. Ilia became professor in Mexico and in 2007 in this capacity invited me to visit his University. We spend marvelous two weeks. He is not less active and energetic as forty years ago. We wish him another forty healthy and productive years to come that would be totally 120, in accord with the Jewish traditional wish.

[2] We write here electrons for concreteness. The formulas presented in this paper are valid for any fast enough charged particle, whose speed $v$ much exceeds the speed of atomic electrons $v_A$, $v \gg v_P$.

We express the inelastic scattering cross-section via characteristics of the target object – the generalized oscillator strength (GOS) density that is the main object of our consideration in this paper.

Until now, photoionization of endohedrals attracted a lot of attention, mainly from the theoretical side [1]. However, recently experimental data became also available (see e.g. [2]). This direction of research is of interest since is able to give valuable information on the structure of a fullerene: atom A inside can serve as a "lamp" that clarifies the $C_N$ structure from the inside, by studying how photoelectrons from A in A@$C_N$ cross the fullerenes shell. Of interest and quite informative is not only the total but also differential in angle photoionization cross-section, mainly its dipole and non-dipole angular anisotropy parameters.

Photoionization supplies the corresponding data as a function of the photon frequency $\omega$. The contribution of the non-dipole terms is suppressed by an additional, as compared to the dipole term, parameter $\omega R/c \ll 1$ [3], where $R$ is the atomic radius and $c$ is the speed of light. Note that if one uses electron spectroscopy technique, the contribution from A in A@$C_N$ can be reliably separated from contribution that comes from $C_N$ electrons

GOS are functions not only of the transferred to atom A in A@$C_N$ energy $\omega$ but also of transferred momentum $q$. In the case of photoionization the relation $\omega = c\kappa$, where $c$ is the speed of light, connects the photon energy $\omega$ and momentum $\kappa$. In fast electron – target scattering similar relation $\omega = \upsilon q$, with $\upsilon \ll c$ being the fast electron speed is not valid any more. Instead, the relation $\omega = \vec{\upsilon}\vec{q} - q^2/2$ holds that transforms into $\omega = \vec{\upsilon}\vec{q}$ for small $q$. It means that fast electron scattering gives dependences of matrix elements of target ionization upon two variables, namely $\omega$ and $q$.

For fast electrons, the main contribution to the cross section comes from the projectile's small scattering angles. It means that for small $q$ that will be the object of our consideration we can consider $\omega \simeq \vec{\upsilon}\vec{q}$. A rather interesting area of research is the investigation of the secondary electrons angular distribution relative to the direction of the transferred momentum $\vec{q}$. In principle, it is similar to investigating the angular distribution of photoelectrons in the photoionization process. The essential difference, however, is that the non-dipole terms are suppressed not by the parameter $\omega R/c \ll 1$ but by a much bigger one $\omega R/\upsilon < 1$. Thus, the contribution of the non-dipole, in first turn, quadrupole terms can be considerably enhanced. In addition, in the fast electron inelastic scattering the monopole transition, entirely forbidden in photoionization, shows up. We will demonstrate that while the fast electron inelastic scattering cross-section is simply proportional to the absolute photoionization cross-sections, this is not true for the angular distributions. Hence, this gives another additional argument to study ionization of endohedrals by fast electron

Objects of our research in this paper are endohedral atoms. They are quite complex multi-electron and multi-atom formations, the direct ab-initio computation of whose properties is almost impossible. Therefore, simplifications and the choice of model-type approaches become inevitable. It means that to check the validity of such approaches the input of corresponding experimental investigations on this or that stage is necessary.

The positioning of an atom A inside a fullerenes shell affects both the shell $C_N$ and the atom A. In principle, a number of possibilities are open, including stretching of $C_N$ and simultaneous compression of the atom A and vice versa. Re-dislocation of electrons from A to the $C_N$ shell or in the opposite direction can accompany the formation of endohedrals. Of importance could be a process, in which stuffing of A inside $C_N$ will lead to ionization of $C_N$.

---

[3] Atomic system of units is used in this paper: electron charge $e$, its mass $m$ and Plank constant $\hbar$ being equal to 1, $e = m = \hbar = 1$

Since in this paper we concentrate on the ionization of $A@C_N$ by fast electrons and on the bases of existing knowledge of the endohedral structure, for considered objects $Ar@C_{60}$ and $Xe@C_{60}$ we will neglect the mutual influence of A and $C_{60}$ upon each other in their ground state. In fact, we will take into account only two most prominent effects. The first is reflection of the secondary electrons, knocked out from A, by the fullerene shell [3]. The second is modification of the field, by which the incoming electron acts upon caged atom due to $C_N$ electron shell polarization by the fast electron [4]. Both effect lead to prominent additional structures in the cross-section ionization by photons and fast electrons, namely to new resonances, cross-section oscillations etc.[5]

Random phase approximation with exchange (RPAE) incorporates both these effects after performing a relatively simple modification as compared to pure atomic case. Note, that RPAE proved to be quite effective in studies of such processes as photoeffect and fast electron inelastic scattering in isolated atoms [6, 7].

In this paper we will pay some attention not only to $A@C_N$ but even to a more exotic system. Namely, we will consider as a target an endohedral of the structure $A@C_{60}@C_{240}$ that presents an endohedral $A@C_{60}$ stuffed inside a giant fullerene $C_{240}$.

We will concentrate here on the angular distribution of secondary electrons relative to the momentum $\vec{q}$, transferred to the target in the collision process. We limit ourselves to so small $q$ that permits to consider only several lower polynomials in the angular distribution of the secondary electrons. This allows studying not only dipole, but also monopole, quadrupole and octupole matrix elements, as functions of both $\omega$ and $q$.

Non-dipole corrections to photoionization of isolated atoms were presented for the first time long ago but due to experimental difficulties were observed and investigated starting only about fifteen-twenty years ago (see [8] and references therein). This permitted to study quadrupole continuous spectrum matrix elements of atomic electrons that in the absolute cross photoionization cross-section are unobservable in the shadow of much bigger dipole contribution. However, the information from photoionization studies does not include $q$-dependences and monopole matrix elements.

Quite long ago the fast charged particle inelastic scattering process was considered as a "synchrotron for poor" [9]. This notion reflects the fact that the fast charge particle inelastic scattering is similar to photoionization, since the dipole contribution mainly determines it. However, contrary to the photoionization case, the ratio "quadrupole-to-dipole" contributions can be much bigger, since instead of $\omega R_A / c \ll 1$ they are determined by $\omega R_A / \upsilon$, where $\upsilon$ is the speed of the projectile and $\omega$ is the transferred energy. Therefore, the quadrupole term in inelastic scattering is relatively much bigger. Deep similarity between photoionization and fast electron scattering brought to a belief that not only the total cross-section, but also the angular anisotropy parameters are either the same of similar. As we demonstrated recently, this is incorrect even in the limit $q \to 0$ [10].

In this paper, following the methodology developed in [10], we investigate the differential cross-section of inelastic scattering of fast electrons upon endohedrals as a function of the angle $\theta$ between the momentum of the emitted in collision process electron and the direction of vector $\vec{q}$. The fast charged particle inelastic scattering cross section is proportional to the so-called generalized oscillator strength (GOS) density. Thus, we study in this paper the GOS density angular distribution as a function of the angle $\theta$.

In understanding the angular distributions, it is decisively important that the ionizing field in collision process is longitudinal, while in photoionization – it is transversal.

We demonstrated that contrary to the case of cross-sections, the angular anisotropy parameters for secondary electrons in fast charge particle-endohedral collisions and in photoionization of the same objects are essentially different.

Let us have in mind that with growth of $q$ the detailed fullerenes structure as a system of carbon atoms became important. The approximation employed in this paper neglects this structure. It means that here we limit ourselves to $q \ll 1/\bar{r}_C$ where $\bar{r}_C$ is the inter-atomic distance at the fullerenes surface.

## 2. Main formulae: general and atomic case

The cross-section of the fast electron inelastic scattering upon an endohedral with ionization of an electron from a subshell with principal quantum number $n$ and angular momentum $l$ can be presented [11, 12] as follows

$$\frac{d^2\sigma_{nl}}{d\omega dq} = \frac{2\sqrt{(E-\omega)}}{\sqrt{E}\omega q^2}\frac{dF_{nl}(q,\omega)}{d\omega}. \tag{1}$$

Here $dF_{nl}(q,\omega)/d\omega$ is the GOS density, differential in the ionized electron energy $\varepsilon = \omega - I_{nl}$, where $I_{nl}$ is the $nl$ subshell ionization potential of atom A in A@C$_N$.

The following formula presents the GOS density differential in both the emission angle and energy of the ionized electron with linear momentum $\vec{k}$ from a subshell $nl$ in one-electron HF approximation:

$$\frac{df_{nl}(q,\omega)}{d\Omega} = \frac{1}{2l+1}\frac{2\omega}{q^2}\sum_{m,s}\left|\langle nlms|\exp(i\vec{q}\vec{r})|\varepsilon \vec{k}s\rangle\right|^2. \tag{2}$$

Here $\vec{q} = \vec{p} - \vec{p}\,'$, with $\vec{p}$ and $\vec{p}\,'$ being the linear moments of the fast incoming and outgoing electrons determined by the initial $E$ and final $E'$ energies as $p = \sqrt{2E}$ and $p' = \sqrt{2E'}$, $\Omega$ is the solid angle of the emitted electron, $m$ is the angular momentum projection, $s$ is the electron spin. Note that $\omega = E - E'$, and $\varepsilon = \omega - I_{nl}$ is the outgoing electron energy.

The values of $\omega$ are limited by the relation $0 \le \omega \le pq - q^2/2$, contrary to the proportionality $\omega = cq$ for the case of photoeffect. In order to consider the projectile as fast, its speed must be much higher than the speed of electrons in the ionized subshell, i.e. $\sqrt{2E} \gg R^{-1}$. The transferred to the atom momentum $q$ is small if $qR \le 1$.

Expanding $\exp(i\vec{q}\vec{r})$ into a sum of products of radial and angular parts, assuming that the fullerenes additional field is spherically symmetric and performing analytic integration over the angular variables, one obtains for GOS in one-electron Hartree-Fock approximation:

$$g_{nl,kl',L}(q) \equiv \int_0^\infty R_{nl}(r)j_L(qr)R_{kl'}(r)r^2 dr, \tag{3}$$

where $j_L(qr)$ are the spherical Bessel functions and $R_{nl(kl')}(r)$ are the radial parts of the HF electron wave functions in the initial (final) states.

We suggest measuring the angular distribution of the emitted electrons relative to the vector $\vec{q}$. It means that the z-axis coincides with the direction of $\vec{q}$ and hence one has to put $\theta_{\vec{q}} = \varphi_{\vec{q}} = 0$ in Eq. (2). Since we have in mind ionization of a particular $nl$ subshell, for simplicity of notation let us introduce the following abbreviations $g_{nl,kl',L}(q) \equiv g_{kl'L}(q)$. Note that due to energy conservation in the fast electron inelastic scattering process $k = \sqrt{2(\omega - I_{nl})}$

One can generalize GOS formulas including inter-electron correlations in the frame of RPAE. We will do it separately for an isolated atom and then modify to include the effect of the $C_N$ shell. GOS in RPAE are obtained substituting $g_{kl'L}(q)$ by modulus $\tilde{G}_{kl'L}(q)$ and the scattering phases $\delta_{l'}$ by $\bar{\delta}_{l'} = \delta_{l'} + \Delta_{l'}$, where the expressions $G_{kl'L}(q) \equiv \tilde{G}_{kl'L}(q)\exp(i\Delta_{l'})$ are solutions of the RPAE set of equations [7, 13]:

$$\langle \varepsilon l' | G^L(\omega,q) | nl \rangle = \langle \varepsilon l' | j_L(qr) | nl \rangle + $$
$$+ \left( \sum_{\varepsilon''l'' \leq F, \varepsilon'''l''' > F} - \sum_{\varepsilon''l'' < F, \varepsilon'''l''' \leq F} \right) \frac{\langle \varepsilon'''l''' | G^L(\omega,q) | \varepsilon''l'' \rangle \langle \varepsilon''l'', \varepsilon l' | U | \varepsilon'''l''', nl \rangle_L}{\omega - \varepsilon_{\varepsilon'''l'''} + \varepsilon_{\varepsilon''l''} + i\eta(1 - 2n_{\varepsilon''l''})}. \quad (4)$$

Here $\leq F (> F)$ denotes summation over occupied (vacant) atomic levels in the target atom. Summation over vacant levels includes integration over continuous spectrum, $n_{\varepsilon l}$ is the Fermi step function that is equal to 1 for $nl \leq F$ occupied and 0 for $nl > F$ vacant states; the Coulomb inter-electron interaction matrix element is defined as $\langle \varepsilon''l'', \varepsilon l' | U | \varepsilon'''l''', nl \rangle_L = \langle \varepsilon''l'', \varepsilon l' | r_<^L / r_>^{L+1} | \varepsilon'''l''', nl \rangle - \langle \varepsilon''l'', \varepsilon l' | r_<^L / r_>^{L+1} | nl, \varepsilon'''l''' \rangle$; $\eta \to +0$. In the latter formula notation of smaller (bigger) radiuses $r_< (r_>)$ of interacting electron coordinates comes from the well-known expansion of the Coulomb inter-electron interaction. The necessary details about solving (4) one can find in [7].

For differential in the secondary electron angle GOS density of $nl$ subshell $dF_{nl}(q,\omega)/d\Omega$ the following relations are valid in RPAE (see also [10]):

$$\frac{dF_{nl}(q,\omega)}{d\Omega} = \sum_{L'L''} \frac{dF_{nl}^{L',L''}(q,\omega)}{d\Omega} = \frac{\omega\pi}{q^2} \sum_{L'L''} (2L'+1)(2L''+1) i^{L'-L''} \times$$
$$\sum_{l'=|L'-l|}^{L'+l} \sum_{l''=|L''-l|}^{L''+l} \tilde{G}_{kl'L'}(\omega,q) \tilde{G}_{kl''L''}(\omega,q) i^{l''-l'} (2l'+1)(2l''+1) e^{i(\bar{\delta}_{l'} - \bar{\delta}_{l''})} \begin{pmatrix} L' & l & l' \\ 0 & 0 & 0 \end{pmatrix} \begin{pmatrix} l'' & l & L'' \\ 0 & 0 & 0 \end{pmatrix}$$
$$\sum_{L=|l'-l''|}^{l'+l''} P_L(\cos\theta)(-1)^{L+l}(2L+1) \begin{pmatrix} l' & L & l'' \\ 0 & 0 & 0 \end{pmatrix} \begin{pmatrix} L & L' & L'' \\ 0 & 0 & 0 \end{pmatrix} \begin{Bmatrix} L & L' & L'' \\ l & l'' & l' \end{Bmatrix}.$$
(5)

We obtained this relation by generalizing (2) to include RPAE corrections and performing required analytical integrations and summations over projection of the electrons angular moment $m$ with the help of *Mathematica* package [14].

Integrating (5) over $d\Omega$ gives partial values of GOS $F_{nl}(q,\omega)$, determined in RPAE by the following expressions:

$$F_{nl}(q,\omega) = \sum_{L'} F_{nl}^{L'}(q,\omega) = \frac{4\omega\pi^2}{q^2} \sum_{L'} (2L'+1) \sum_{l'=|L'-l|}^{L'+l} [\tilde{G}_{kl'L'}(\omega,q)]^2 (2l'+1) \begin{pmatrix} L' & l & l' \\ 0 & 0 & 0 \end{pmatrix}^2. \quad (6)$$

Note that at small $q$ the dipole contribution in GOSes $F_{nl}(q,\omega)$ dominates and is simply proportional to the photoionization cross-section $\sigma_{nl}(\omega)$ [12]. To compare the results obtained with known formulas for the photoionization with the lowest order of non-dipole corrections taken into account, let us consider so small $q$ that it is enough to take into account terms with $L',L'' \leq 2$. In this case, one can present GOS angular distribution (5) similarly to the photoionization case (see e.g. [8, 13])

$$\frac{dF_{nl}(q,\omega)}{d\Omega} = \frac{F_{nl}(q,\omega)}{4\pi}\left\{1 - \frac{\beta_{nl}^{(in)}(\omega,q)}{2}P_2(\cos\theta) + q\left[\gamma_{nl}^{(in)}(\omega,q)P_1(\cos\theta) + \eta_{nl}^{(in)}(\omega,q)P_3(\cos\theta) + \varsigma_{nl}^{(in)}(\omega,q)P_4(\cos\theta)\right]\right\}.$$

(7)

The obvious difference is the $q$ dependence of the coefficients and an extra term $\varsigma_{nl}^{(in)}(\omega,q)P_4(\cos\theta)$. Even in this case, expressions for $\beta_{nl}^{(in)}(\omega,q)$, $\gamma_{nl}^{(in)}(\omega,q)$, $\eta_{nl}^{(in)}(\omega,q)$, and $\varsigma_{nl}^{(in)}(\omega,q)$ via $g_{kl'L'}(q)$ are too complex as compared to relations for $\beta_{nl}(\omega)$, $\gamma_{nl}(\omega)$, and $\eta_{nl}(\omega)$ in photoionization [13]. Therefore, it is more convenient to present the results for $s$, $p$, and $d$ subshells separately. We demonstrate that while $F_{nl}(q,\omega) \sim \sigma(\omega)$, similar relations are not valid for the anisotropy parameters.

The calculations of GOS in this paper is performed in the frame of the random phase approximation with exchange (RPAE) modified in a way that permits to include the static effect of the $C_{60}$ shell via a zero-thickness potential [15] and expresses the dynamic action of $C_{60}$ upon the caged atoms via fullerenes dipole polarizability [4]. The credibility of RPAE is well established for atoms [6], while the reasonable accuracy of the zero-thickness potential at least in obtaining qualitative predictions is confirmed by comparing the calculation [16] and measured [17] data for 4d subshell of Xe@$C_{60}$.

### 3. Main formulae: endohedrals

We denote $G_{kl'L}^{AC}(\omega,q)$ the GOS amplitude of an electron transition from an endohedral atom's $nl$ subshell into continuum state, characterized by linear momentum $k$, energy $\varepsilon = k^2/2$, connected to the transferred to the endohedral energy $\omega$ by the relation $\varepsilon = \omega - I_{nl}$, where $I_{nl}$ is the $nl$ subshell ionization potential, and angular momentum $l'$. The amplitude $G_{kl'}^{AC}(\omega,q)$ takes into account in RPAE frame the reflection of photoelectrons by the $C_{60}$ shell and polarization of the latter under the action of the incoming photon beam. Remarkable that for fast enough projectiles the amplitude is the following product [18]

$$G_{\varepsilon l',L}^{AC}(\omega,q) = G_L(\omega,q)F_{l'}(k)G_{\varepsilon l',L}^{F}(\omega,q).$$

(8)

The polarization factor $G_L(\omega,q)$ takes into account the modification of the interaction between the incoming fast particle and endohedral electron due to presence of the fullerene $C_N$; $F_{l'}(k)$ describes the reflection factor that represents the effect of the fullerenes shell $C_N$ upon the knocked out from atom A secondary electrons with the angular momentum $l'$. In (8), $G_{\varepsilon l', L}^F(\omega, q)$ is the atomic ionization amplitude, in which the virtual states are modified due to action of the static potential of the fullerenes shell upon the virtually excited atomic states.

According to our estimates, polarization most prominently acts in the dipole channel, $L = 1$. In the calculations performed in the current paper, we take into account corrections to the dipole channel only and neglect the effect of polarization in other channels putting in calculations $G_L(\omega, q) = 1$ for all $L \neq 1$. Since the radius $R_A$ of the atom A is considerably smaller than the fullerenes radius $R_C$, the expression for $G_1(\omega, q)$ becomes rather simple

$$G_1(\omega, q) = 1 - \frac{\alpha_C(\omega, q)}{R_C^2} . \qquad (9)$$

where $\alpha_C(\omega, q)$ is the generalized dipole polarizability of a fullerene. Since we are interested in small $q$, we can substitute in (9) the generalized polarizability by the ordinary one $\alpha_C(\omega) \equiv \alpha_C(\omega, q)|_{q \to 0}$.

The dipole polarizability $\alpha_C(\omega)$ is difficult to calculate ab-initio, but one can easily expressed (see [4] and references therein) via experimentally quite well known photoionization cross-section $\sigma_C(\omega)$ of the C$_{60}$ [19]:

$$\text{Re}\,\alpha_C(\omega) = \frac{c}{2\pi^2} \int_{I_F}^{\infty} \frac{\sigma_C(\omega')d\omega'}{\omega'^2 - \omega^2}, \text{Im}\,\alpha_C(\omega) = c\sigma_C(\omega)/4\pi\omega . \qquad (10)$$

Here $I_C$ is the fullerene ionization potential and $c$ is the speed of light.

Since the cross-section $\sigma_C(\omega)$ is absolutely dominated by the fullerenes Giant resonance that have a maximum at about 2Ry, $G(\omega)$ starts to decreases rapidly at $\omega > 2Ry$ reaching its asymptotic value equal to 1 at about 5Ry. Note that this factor, connecting the atomic and fullerenes GOS, is able to alter the endohedral generalized oscillator strength as compared to corresponding pure atomic value.

To obtain the reflection factor $F_{l'}(k)$, we substitute the fullerenes shell action by a static zero-thickness potential [15]

$$W(r) = -W_0 \delta(r - R_C) . \qquad (11)$$

We obtain the parameter $W_0$ from the condition that the binding energy of extra electron in negative ion $C_{60}^-$ is equal to the experimentally observed value. The factor $F_{l'}(k)$ is determined by the expression (see [10, 15] and references therein):

$$F_{l'}(k) = \cos\Delta\delta_{l'}(k)\left[1 - \tan\Delta\delta_{l'}(k)\frac{v_{kl'}(R)}{u_{kl'}(R)}\right] = \frac{k\cos\Delta\delta_{l'}(k)}{k - 2W_0 u_{kl'}(R_C) v_{kl'}(R_C)}, \qquad (12)$$

where $\Delta\delta_{l'}(k)$ is the addition elastic scattering phase of knocked-out electron partial wave $l'$ due to action of the potential (11), $u_{kl'}(r)$ is the regular and $v_{kl'}(r)$ irregular at point $r \to 0$ radial parts of atomic Hartree-Fock one-electron wave functions. The following relation expresses the additional phase shift $\Delta\delta_{l'}(k)$ [10, 15]:

$$\tan \Delta \delta_{l'}(k) = \frac{u_{kl'}^2(R_C)}{u_{kl'}(R_C)v_{kl'}(R_C) - k/2W_0} .\qquad(13)$$

The factor $F_{l'}(k)$ as a function of $k$ oscillates due to interference between the direct photoelectron wave and its reflections from the fullerenes shell. This factor redistributes the resulting GOS as compared to that of the isolated atom but cannot change their value integrated over essential $\omega$ region.

We obtain $G^F_{\varepsilon l',L}(\omega,q)$ in the frame of the RPAE. When potential (11) presents the fullerenes shell, the following equation [10, 18] is valid:

$$\begin{aligned}\langle \varepsilon l' | G^F_L(\omega,q) | nl \rangle &= \langle \varepsilon l' | j_L(qr) | nl \rangle + \\ &+ \sum_{\varepsilon''l'',\varepsilon'''l'''} \frac{\langle \varepsilon'''l''' | G^F_L(\omega,q) | \varepsilon''l'' \rangle \left[ F^2_{\varepsilon''l''} n_{\varepsilon'''l'''}(1-n_{\varepsilon''l''}) - F^2_{\varepsilon'''l'''} n_{\varepsilon''l''}(1-n_{\varepsilon'''l'''}) \right]}{\omega - \varepsilon_{\varepsilon''l''} + \varepsilon_{\varepsilon'''l'''} + i\eta(1-2n_{\varepsilon''l''})} \times \\ &\times \langle \varepsilon''l'', \varepsilon l' | U | \varepsilon'''l''', nl \rangle_L .\end{aligned}\qquad(14)$$

Here $\langle \varepsilon l' | G^F_L(\omega,q) | nl \rangle \equiv G^F_{l'L}(\omega,q) = \tilde{G}^F_{l'L}(\omega,q)\exp(i\overline{\Delta}_{l'})$, $F_{\varepsilon' l'} \equiv F_{l'}(k')$, $j_L(qr)$ is the one-electron operator that describes photon-electron interaction. The necessary details about solving (14) one can find in [7, 10].

Let we start with s-subshells, where the following relation gives differential GOSes in the above-mentioned $L',L'' \leq 2$ approximation. Note that we omit for simplicity of the expressions that follow the upper index $F$ and the arguments $\omega,q$ from $\tilde{G}^F_{l'L}(\omega,q)$ reducing it to $\tilde{G}_{l'L}$

$$\begin{aligned}\frac{dF_{ns}(q,\omega)}{d\Omega} &= \sum_{L',L''=0}^{2} \frac{dF^{L',L''}_{ns}(q,\omega)}{d\Omega} = \frac{F_{ns}(q,\omega)}{4\pi}\Big\{ 1 + \frac{6}{W_0}\tilde{G}_{11}\left[ \tilde{G}_{00}\cos(\overline{\delta}_0 - \overline{\delta}_1) + 2\tilde{G}_{22}\cos(\overline{\delta}_1 - \overline{\delta}_2) \right] \\ &P_1(\cos\theta) + \frac{2}{7W_0}\left[ 21\tilde{G}^2_{11} + 5\tilde{G}_{22}(7\tilde{G}_{00}\cos(\overline{\delta}_0 - \overline{\delta}_2) + 5\tilde{G}_{22}) \right] P_2(\cos\theta) + \\ &\frac{18}{W_0}\tilde{G}_{11}\tilde{G}_{22}\cos(\overline{\delta}_1 - \overline{\delta}_2) P_3(\cos\theta) + \frac{90}{7W_0}\tilde{G}^2_{22} P_4(\cos\theta) \Big\} \\ &\equiv \frac{F_{ns}(q,\omega)}{4\pi}\Big\{ 1 - \frac{1}{2}\beta^{(in)}_{ns}(q,\omega)P_2(\cos\theta) + q\left[ \gamma^{(in)}_{ns}(q,\omega)P_1(\cos\theta) + \eta^{(in)}_{ns}(q,\omega)P_3(\cos\theta) \right] \Big\}.\end{aligned}\qquad(15)$$

Here

$$F_{ns} = \frac{4\pi^2\omega}{q^2}W_0;\ W_0 = \tilde{G}^2_{00} + 3\tilde{G}^2_{11} + 5\tilde{G}^2_{22}.\qquad(16)$$

The expressions (15, 16), and corresponding expressions for higher $l$, look similar to that for isolated atoms [10]. The difference here is in that the amplitude $G_{l'L}$ we determine by (8), with upper indexes AC omitted for compactness of expressions. In spite of the similarity, we find it necessary and convenient for the reader to present them here for clarification of the analyses.

We will compare the result obtained in the small $q$ limit with the known formula for photoionization of an atom by non-polarized light [13]. To do this, we have to use the lowest order terms of the first three spherical Bessel functions:

$$j_0(qr) \cong 1 - \frac{(qr)^2}{6}; \quad j_1(qr) \cong \frac{qr}{3}\left(1 - \frac{(qr)^2}{10}\right); \quad j_2(qr) \cong \frac{(qr)^2}{15}\left(1 - \frac{(qr)^2}{14}\right); \quad j_3(qr) \cong \frac{(qr)^3}{105}. \quad (17)$$

The lowest in powers of $q$ term is $\tilde{G}_{11} \sim q \ll 1$[4]. Correction to $\tilde{G}_{11}$ is proportional to $q^3$. As to $\tilde{G}_{00}$ and $\tilde{G}_{22}$, they are proportional to $q^2$ with additions of the order of $q^4$. By retaining in (15) terms of the order of $q^2$ and bigger, one obtains the following expression:

$$\frac{dF_{ns}(q,\omega)}{d\Omega} = \frac{F_{ns}(q,\omega)}{4\pi}\{1 + 2P_2(\cos\theta) +$$
$$\frac{2}{\tilde{G}_{11}}\left[\tilde{G}_{00}\cos(\bar{\delta}_0 - \bar{\delta}_1) + 2\tilde{G}_{22}\cos(\bar{\delta}_1 - \bar{\delta}_2)\right]P_1(\cos\theta) + \frac{6\tilde{G}_{22}}{\tilde{G}_{11}}\cos(\bar{\delta}_1 - \bar{\delta}_2)P_3(\cos\theta)\} \equiv \quad . \quad (18)$$
$$\equiv \frac{F_{ns}(q,\omega)}{4\pi}\{1 + 2P_2(\cos\theta) + q\left[\gamma_{ns}^{(in)}(q,\omega)P_1(\cos\theta) + \eta_{ns}^{(in)}(q,\omega)P_3(\cos\theta)\right]\}$$

One should compare this relation with the similar one for photoionization of $ns$ subshell [13]:

$$\frac{d\sigma_{ns}(\omega)}{d\Omega} = \frac{\sigma_{ns}(\omega)}{4\pi}\{1 - P_2(\cos\theta) + \kappa[\gamma_{ns}(\omega)P_1(\cos\theta) + \eta_{ns}(\omega)P_3(\cos\theta)]\}., \quad (19)$$

where $\kappa = \omega/c$ is the photon momentum, and $\gamma_{ns}(\omega) = -\eta_{ns}(\omega) = [6\tilde{Q}_2/5\tilde{D}_1]\cos(\bar{\delta}_1 - \bar{\delta}_2)$; $\tilde{D}_1$ and $\tilde{Q}_2$ are the dipole and quadrupole RPAE matrix elements of the endohedrals photoionization [13].

We see the difference between (18) and (19) in the sign and magnitude of the dipole parameter. This parameter in electron scattering is two times bigger than in photoionization and of opposite sign. Essentially different are expressions for the non-dipole terms. This difference exists and remains even in the so-called optical limit $q \to 0$.

According to (17), in the $q \to 0$ limit there are simple relations between dipole $\tilde{D}_1$ and quadrupole $\tilde{Q}_2$ matrix elements and functions $\tilde{G}_{11}$, $\tilde{G}_{22}$, namely $\tilde{G}_{11} = q\tilde{D}_1/3$ and $\tilde{G}_{22} = 2q^2\tilde{Q}_2/15$. With the help of these relations and $\tilde{G}_{00} = -q^2\tilde{Q}_2/3 = -(5/2)\tilde{G}_{22}$, (18) transforms into the following expression:

$$\frac{dF_{ns}(q,\omega)}{d\Omega} = \frac{F_{ns}(q,\omega)}{4\pi} \times$$
$$\left\{1 + 2P_2(\cos\theta) + q\left[\frac{2\tilde{Q}_2}{\tilde{D}_1}\left(\frac{4}{5}\cos(\bar{\delta}_1 - \bar{\delta}_2) - \cos(\bar{\delta}_0 - \bar{\delta}_1)\right)P_1(\cos\theta) + 2\gamma_{ns}(\omega)P_3(\cos\theta)\right]\right\}. \quad (20)$$

---

[4] As is seen from (17), we have in mind such values of $q$ that inequality $qR < 1$ is valid.

The deviation from (19) is evident, since one cannot express the angular distribution in inelastic scattering via a single non-dipole parameter $\gamma_{ns}(\omega)$ including an absent in photoionization phase difference $\bar{\delta}_0 - \bar{\delta}_1$. As a result, the following relations have to be valid at very small $q$:

$$\gamma_{ns}^{(in)}(\omega) = \frac{2\tilde{Q}_2}{\tilde{D}_1}\left[\frac{4}{5}\cos(\bar{\delta}_1 - \bar{\delta}_2) - \cos(\bar{\delta}_0 - \bar{\delta}_1)\right]; \; \eta_{ns}^{(in)}(\omega) = 2\gamma_{ns}(\omega) = \frac{12}{5}\frac{\tilde{Q}_2}{\tilde{D}_1}\cos(\bar{\delta}_1 - \bar{\delta}_2)., \quad (21)$$

We see that the investigation of inelastic scattering even at $q \to 0$ permits to obtain an additional characteristic of the ionization process, namely, its s-wave phase.

4. **Angular anisotropy parameters for *p* and *d*-subshells**

We demonstrate below that for $l > 0$, even at very small $q$, the relations between non-dipole parameters in photoionization and inelastic fast electron scattering are more complex.

The similarity of general structure and considerable difference between (18) and (19) is evident. Indeed, the contribution of the non-dipole parameters one can enhance, since the condition $\omega/c \ll q \ll R_A^{-1}$ is easy to achieve. Let us note that even while neglecting the terms with $q$, (19) and (20) remain different: in photoionization, the angular distribution is proportional to $\sin^2\theta$ (see (19)), whereas in inelastic scattering it is proportional to $\cos^2\theta$ (see (20)). The reason for this difference is clear. In photoabsorption, the atomic electron is "pushed" off the endohedral by the electric field of the photon, which is perpendicular to the direction of the light beam. In inelastic scattering, the "push" acts along momentum $\vec{q}$, so the preferential emission of the electrons takes place along the $\vec{q}$ direction, so the maximum is at $\theta = 0$. Similar reason explains the difference in the non-dipole terms. Note that the last term due to monopole transition in (20) is absent in photoabsorption angular distribution (19). It confirms that the angular distribution of the GOS densities is richer than that of photoionization.

Although the expressions for *p*- and *d*-subshells are much more complex that for s, they are of great importance and interest since *p*- and *d*– subshells are multi-electron objects. Intra-shell electron correlations affect the ionization of these subshells quite strong. Particularly important are the multi-electron effects in 4*d*- subshell due to presence there of the famous dipole Giant resonance (see e.g. [6]). This is why it is of interest to present data on non-s-subshells also.

We obtain the following expression for differential GOS of *p*-subshells ($l = 1$):

$$\frac{dF_{np}(q,\omega)}{d\Omega} = \sum_{L',L''=0}^{2} \frac{dF_{np}^{L',L''}(q,\omega)}{d\Omega} = \frac{F_{np}}{4\pi}\{1+$$

$$\frac{1}{5W_1}\Big[10\tilde{G}_{01}(2\tilde{G}_{12}-\tilde{G}_{10})\cos(\bar{\delta}_0-\bar{\delta}_1)+4\tilde{G}_{21}\big((5\tilde{G}_{10}-\tilde{G}_{12})\cos(\bar{\delta}_1-\bar{\delta}_2)+9\bar{G}_{32}\cos(\bar{\delta}_2-\bar{\delta}_3)\big)\Big]P_1(\cos\theta)+$$

$$\frac{2}{7W_1}\Big[7\tilde{G}_{21}(\tilde{G}_{21}-2\tilde{G}_{01}\cos(\bar{\delta}_0-\bar{\delta}_2))+7\tilde{G}_{12}(\tilde{G}_{12}-2\tilde{G}_{10})+$$

$$3\tilde{G}_{32}((7\tilde{G}_{10}-2\tilde{G}_{12})\cos(\bar{\delta}_1-\bar{\delta}_3)+4\tilde{G}_{32})\Big]P_2(\cos\theta)-$$

$$\frac{6}{5W_1}\Big[6\tilde{G}_{21}\tilde{G}_{12}\cos(\bar{\delta}_1-\bar{\delta}_2)+\tilde{G}_{32}(5\tilde{G}_{01}\cos(\bar{\delta}_0-\bar{\delta}_3)-4\tilde{G}_{21}\cos(\bar{\delta}_2-\bar{\delta}_3))\Big]P_3(\cos\theta)+$$

$$\frac{18}{7W_1}\tilde{G}_{32}\Big[\tilde{G}_{32}-4\tilde{G}_{12}\cos(\bar{\delta}_1-\bar{\delta}_3)\Big]P_4(\cos\theta)\} \equiv \frac{F_{np}(q,\omega)}{4\pi}\times$$

$$\left\{1-\frac{\beta_{np}^{(in)}(q,\omega)}{2}P_2(\cos\theta)+q\Big[\gamma_{np}^{(in)}(q,\omega)P_1(\cos\theta)+\eta_{np}^{(in)}(q,\omega)P_3(\cos\theta)+\zeta_{np}^{(in)}(q,\omega)P_4(\cos\theta)\Big]\right\},$$

(22)

where

$$F_{np} = \frac{4\pi^2\omega}{q^2}W_1;\quad W_1 = \tilde{G}_{10}^2+\tilde{G}_{01}^2+2\Big[\tilde{G}_{21}^2+\tilde{G}_{12}^2\Big]+3\tilde{G}_{32}^2. \quad (23)$$

For differential GOSes of *d*-subshells ($l=2$) the following expression holds:

$$\frac{dF_{nd}(q,\omega)}{d\Omega} = \frac{F_{nd}}{4\pi}\{1+\frac{6}{W_2}\Big[14\tilde{G}_{11}(\tilde{G}_{22}-\tilde{G}_{20})\cos(\bar{\delta}_1-\bar{\delta}_2)-14\tilde{G}_{11}\tilde{G}_{02}\cos(\bar{\delta}_0-\bar{\delta}_1)+$$

$$3\tilde{G}_{31}\big((7\tilde{G}_{20}-2\tilde{G}_{22})\cos(\bar{\delta}_2-\bar{\delta}_3)+12\tilde{G}_{42}\cos(\bar{\delta}_3-\bar{\delta}_4)\big)\Big]P_1(\cos\theta)+$$

$$\frac{2}{245W_2}\Big[1029(\tilde{G}_{11}^2+6\tilde{G}_{31}^2)-18522\tilde{G}_{11}\tilde{G}_{31}\cos(\bar{\delta}_1-\bar{\delta}_3)+1225\tilde{G}_{02}(7\tilde{G}_{20}-10\tilde{G}_{22})\cos(\bar{\delta}_0-\bar{\delta}_2)-$$

$$125\tilde{G}_{22}(98\tilde{G}_{20}+15\tilde{G}_{22})+450\tilde{G}_{42}((49\tilde{G}_{20}-20\tilde{G}_{22})\cos(\bar{\delta}_2-\bar{\delta}_4)+25\tilde{G}_{42})\Big]P_2(\cos\theta)+$$

$$\frac{18}{W_2}\Big[2\tilde{G}_{11}(\tilde{G}_{22}\cos(\bar{\delta}_1-\bar{\delta}_2)-6\tilde{G}_{42}\cos(\bar{\delta}_1-\bar{\delta}_4))+\tilde{G}_{31}(7\tilde{G}_{02}\cos(\bar{\delta}_0-\bar{\delta}_3)-$$

$$8\tilde{G}_{22}\cos(\bar{\delta}_2-\bar{\delta}_3)+6\tilde{G}_{42}\cos(\bar{\delta}_3-\bar{\delta}_4))\Big]P_3(\cos\theta)+$$

$$\frac{90}{49W_2}\Big[20\tilde{G}_{22}^2+\tilde{G}_{42}(98\tilde{G}_{02}\cos(\bar{\delta}_0-\bar{\delta}_4)-100\tilde{G}_{22}\cos(\bar{\delta}_2-\bar{\delta}_4)+27\tilde{G}_{42})\Big]P_4(\cos\theta)\} \equiv$$

$$\frac{F_{nd}(q,\omega)}{4\pi}\times$$

$$\left\{1-\frac{\beta_{nd}^{(in)}(q,\omega)}{2}P_2(\cos\theta)+q\Big[\gamma_{nd}^{(in)}(q,\omega)P_1(\cos\theta)+\eta_{nd}^{(in)}(q,\omega)P_3(\cos\theta)+\zeta_{nd}^{(in)}(q,\omega)P_4(\cos\theta)\Big]\right\},$$

(24)

where

$$F_{nd} = \frac{4\pi^2 \omega}{35 q^2} W_2; \quad W_2 = 35\tilde{G}_{20}^2 + 42\tilde{G}_{11}^2 + 63\tilde{G}_{31}^2 + 35\tilde{G}_{02}^2 + 50\tilde{G}_{22}^2 + 90\tilde{G}_{42}^2. \tag{25}$$

It is interesting to compare, just as was done with $l = 0$, the expressions (22) and (24) with angular distribution of photoelectrons. It is essential to clarify whether the difference exists even in the $q \to 0$ limit, as it takes place for the $s$- subshells. In this limit, the following expressions follow from (22) and (24):

For $l = 1$ one has from (22) at $q = 0$

$$\beta_{np}^{(in)}(q=0, \omega) = -\frac{4}{\tilde{D}_0^2 + 2\tilde{D}_2^2}[\tilde{D}_2^2 - 2\tilde{D}_0 \tilde{D}_2 \cos(\bar{\delta}_0 - \bar{\delta}_2)], \tag{26}$$

$$\gamma_{np}^{(in)}(q=0, \omega) = \frac{18}{25[\tilde{D}_0^2 + 2\tilde{D}_2^2]} \{5\tilde{D}_0 \tilde{Q}_1 \cos(\bar{\delta}_1 - \bar{\delta}_0) + 2\tilde{D}_2[2\tilde{Q}_3 \cos(\bar{\delta}_3 - \bar{\delta}_2) - 3\tilde{Q}_1 \cos(\bar{\delta}_1 - \bar{\delta}_2)]\}, \tag{27}$$

$$\eta_{np}^{(in)}(q=0, \omega) = \frac{12}{25[\tilde{D}_0^2 + 2\tilde{D}_2^2]} \{5\tilde{D}_0 \tilde{Q}_3 \cos(\bar{\delta}_3 - \bar{\delta}_0) + 2\tilde{D}_2 \left[3\tilde{Q}_1 \cos(\bar{\delta}_1 - \bar{\delta}_2) - 2\tilde{Q}_3 \cos(\bar{\delta}_3 - \bar{\delta}_2)\right]\}. \tag{28}$$

For $l = 2$ one has from (24) at $q = 0$

$$\beta_{nd}^{(in)}(q=0, \omega) = -\frac{4}{5[2\tilde{D}_1^2 + 3\tilde{D}_3^2]}[\tilde{D}_1^2 + 6\tilde{D}_3^2 - 18\tilde{D}_1 \tilde{D}_3 \cos(\bar{\delta}_1 - \bar{\delta}_3)], \tag{29}$$

$$\gamma_{nd}^{(in)}(q=0, \omega) = \frac{2}{35[2\tilde{D}_1^2 + 3\tilde{D}_3^2]} \times \{14\tilde{D}_1 \left[7\tilde{Q}_2 \cos(\bar{\delta}_1 - \bar{\delta}_2) - 2\tilde{Q}_0 \cos(\bar{\delta}_0 - \bar{\delta}_1)\right] + 9\tilde{D}_3 \left[8\tilde{Q}_4 \cos(\bar{\delta}_4 - \bar{\delta}_3) - 13\tilde{Q}_2 \cos(\bar{\delta}_2 - \bar{\delta}_3)\right]\}, \tag{30}$$

$$\eta_{nd}^{(in)}(q=0, \omega) = \frac{12}{35[2\tilde{D}_1^2 + 3\tilde{D}_3^2]} \{2\tilde{D}_1 \left[\tilde{Q}_2 \cos(\bar{\delta}_2 - \bar{\delta}_1) - 6\tilde{Q}_4 \cos(\bar{\delta}_4 - \bar{\delta}_1)\right] + \tilde{D}_3 \left[7\tilde{Q}_0 \cos(\bar{\delta}_0 - \bar{\delta}_3) - 8\tilde{Q}_2 \cos(\bar{\delta}_2 - \bar{\delta}_3) - 6\tilde{Q}_4 \cos(\bar{\delta}_4 - \bar{\delta}_3)\right]\}. \tag{31}$$

We use the following relations in deriving formulas (26-31)

$$\tilde{G}_{l'1} \equiv \frac{q}{3}\tilde{D}_{l'}(l' = l \pm 1); \tilde{G}_{l'0} \equiv -\frac{q^2}{3}\tilde{Q}_{l'}(l' = l); \tilde{G}_{l'2} \equiv \frac{2q^2}{15}\tilde{Q}_{l'}(l' = l, l \pm 2); \tag{32}$$

To prevent misunderstanding in notation, let us present the relations that we use in the HF approximation for dipole and quadrupole radial matrix elements $\tilde{d}_{l'}$ and $\tilde{q}_{l'}$:

$$\tilde{D}_{l'} \Rightarrow \tilde{d}_{l'} = \int_0^\infty P_{nl}(r) r P_{kl'}(r) dr; \quad \tilde{Q}_{l'} \Rightarrow \tilde{q}_{l'} = \frac{1}{2}\int_0^\infty P_{nl}(r) r^2 P_{kl'}(r) dr, \tag{33}$$

where $P_{nl(kl')}(r) = rR_{nl(kl')}(r)$ and $R_{nl(kl')}(r)$ are the radial parts of the HF electron wave functions in the initial (final) states.

For any $l$, the angular distribution of photoelectrons with inclusion of non-dipole terms in the lowest order of photon momentum $\kappa$ is given by the following expression:

$$\frac{d\sigma_{nl}(\omega)}{d\Omega} = \frac{\sigma_{nl}(\omega)}{4\pi}\left\{1 - \frac{\beta_{nl}(\omega)}{2}P_2(\cos\theta) + \kappa\gamma_{nl}(\omega)P_1(\cos\theta) + \kappa\eta_{nl}(\omega)P_3(\cos\theta)\right\}. \quad (34)$$

For $l=1$ one has the following expression for the dipole angular anisotropy parameters [6]

$$\beta_{np}(\omega) = \frac{2}{\tilde{D}_0^2 + 2\tilde{D}_2^2}[\tilde{D}_2^2 - 2\tilde{D}_0\tilde{D}_2\cos(\bar{\delta}_0 - \bar{\delta}_2)]. \quad (35)$$

As it is seen from (26), for $l=1$ the relation $\beta_{np}^{(in)}(q=0,\omega) = -2\beta_{np}(\omega)$ is the same as for the $s$-subshells.

The following expressions determine the non-dipole angular anisotropy parameters [8] for $l=1$:

$$\gamma_{np}(\omega) = \frac{6}{25[\tilde{D}_0^2 + 2\tilde{D}_2^2]}\left\{5\tilde{D}_0\tilde{Q}_1\cos(\bar{\delta}_1 - \bar{\delta}_0) + \tilde{D}_2\left[9\tilde{Q}_3\cos(\bar{\delta}_3 - \bar{\delta}_2) - \tilde{Q}_1\cos(\bar{\delta}_1 - \bar{\delta}_2)\right]\right\}, \quad (36)$$

$$\eta_{np}(\omega) = \frac{6}{25[\tilde{D}_0^2 + 2\tilde{D}_2^2]}\left\{5\tilde{D}_0\tilde{Q}_3\cos(\bar{\delta}_3 - \bar{\delta}_0) + 2\tilde{D}_2\left[3\tilde{Q}_1\cos(\bar{\delta}_1 - \bar{\delta}_2) - 2\tilde{Q}_3\cos(\bar{\delta}_3 - \bar{\delta}_2)\right]\right\}. \quad (37)$$

One has for the dipole angular anisotropy parameter the following expression [6] for $l=2$:

$$\beta_{nd}(\omega) = \frac{2}{5[2\tilde{D}_1^2 + 3\tilde{D}_3^2]}[\tilde{D}_1^2 + 6\tilde{D}_3^2 - 18\tilde{D}_1\tilde{D}_3\cos(\bar{\delta}_1 - \bar{\delta}_3)]. \quad (38)$$

Note that for $l=2$, as it is seen from (22, 29 29, 36), the relation similar to $l=0;1$ is valid, namely $\beta_{nd}^{(in)}(q=0,\omega) = -2\beta_{nd}(\omega)$. Quite possible that such a relation is valid for any $l$.

The following expressions determine the non-dipole angular anisotropy parameters [8] for $l=2$

$$\gamma_{nd}(\omega) = \frac{6}{35[2\tilde{D}_1^2 + 3\tilde{D}_3^2]}\left\{7\tilde{D}_1[\tilde{Q}_2\cos(\bar{\delta}_2 - \bar{\delta}_1) - \tilde{Q}_0\cos(\bar{\delta}_0 - \bar{\delta}_1)] + \right.$$
$$\left. 3\tilde{D}_3[6\tilde{Q}_4\cos(\bar{\delta}_4 - \bar{\delta}_3) - \tilde{Q}_2\cos(\bar{\delta}_2 - \bar{\delta}_3)]\right\}, \quad (39)$$

$$\eta_{nd}(\omega) = \frac{6}{35[2\tilde{D}_1^2 + 3\tilde{D}_3^2]}\left\{2\tilde{D}_1\left[6\tilde{Q}_4\cos(\bar{\delta}_4 - \bar{\delta}_1) - \tilde{Q}_2\cos(\bar{\delta}_2 - \bar{\delta}_1)\right] - \right.$$
$$\left. \tilde{D}_3\left[8\tilde{Q}_2\cos(\bar{\delta}_2 - \bar{\delta}_3) - 6\tilde{Q}_4\cos(\bar{\delta}_4 - \bar{\delta}_3) - 7\tilde{Q}_0\cos(\bar{\delta}_0 - \bar{\delta}_3)\right]\right\}. \quad (40)$$

Prominent analytic deviation from respective non-dipole parameters for inelastic scattering, given by (20, 21, 22, 24) is clearly seen. Contrary to the dipole parameters, simple frequency independent relations that connect respective non-dipole parameters for photoionization and fast electron inelastic scattering do not exist.

Note that the limit $q = 0$ at $\omega \neq 0$ cannot be achieved since no energy can be transferred from the incoming electron to the projectile without momentum transfer. However, with growth of the projectile's speed, smaller and smaller $q$ is sufficient to transfer the given energy $\omega$.

We note that in spite of visibly deep similarity between photoionization and fast electron scattering, a big difference exists. Indeed, the angular distributions in photoionization and fast electron scattering are different even in the limit $q \to 0$. One can explain it by the difference between a transverse (in photoionization) and longitudinal (in fast electron scattering) photons that ionize the target atom. We see it analytically, in the difference between operators causing ionization by photons and fast electrons that include already only the lowest non-dipole corrections. For photoionization, this is the operator $(\vec{e}\vec{r}) + i(\vec{\kappa}\vec{r})(\vec{e}\vec{r})$, where $\vec{e}$ is the photon polarization vector that is orthogonal to the direction of light propagation given by the photon momentum $\vec{\kappa}$. As to fast electron scattering, it is $(\vec{q}\vec{r}) + i(\vec{q}\vec{r})(\vec{q}\vec{r})$ that includes only one angle between $\vec{q}$ and $\vec{r}$ contrary to the case of photoionization with its two angles – between $\vec{r}$, $\vec{e}$ and $\vec{\kappa}$.

Because of this difference, in photoionization the force that acts upon the outgoing electron is orthogonal to the direction of $\vec{\kappa}$ and thus of the photon beam. Therefore, the photoelectron emission is minimal along $\vec{\kappa}/\kappa$, while in inelastic electron scattering the force and maximal knocked-out electron yield is directed along $\vec{q}$.

## 4. Onion-type fullerenes

Of additional interest are the recently discovered [20] onion-type endohedrals. In these objects inside a very big fullerene $C_{N_2}$, of which concrete example is $C_{240}$, which can absorb an entire endohedral $A@C_{N_1}$, e.g. $Ar@C_{60}$. We denote such a molecule as $A@C_{N_1}@C_{N_2}$. Let us assume, on the ground of existing data that the outer fullerenes radius $R_{N_2}$ is much bigger than $R_{N_1}$ and the strong inequality $R_{N_2} \gg R_{N_1} \gg R_A$ is valid. In fact, as we will see in this Section the strong inequality is not that strong. However, the accepting of this inequality permits to express the action of two fullerenes shells as combination of parameters that characterize each of the fullerenes shell separately.

$$W(r) = -W_1 \delta(r - R_{f1}) - W_2 \delta(r - R_{f2}) \tag{41}$$

According to [21]

$$F_{l'}(\omega) = \frac{k \sin \Delta \delta_{l'}^R}{2\left[W_1 u_1^2 + W_2 u_2^2 - 2W_1 W_2 u_1 u_2 (u_1 v_2 - u_2 v_1)/k\right]} \tag{42}$$

$u_{kl'}(r)$ is the regular and $v_{kl'}(r)$ irregular at point $r \to 0$ radial parts of atomic Hartree-Fock one-electron wave functions.

$$\tan \Delta \delta_{l'}^R(k) = \frac{u_1^2 + u_2 W_2 \left[u_2/W_1 + 2u_1 W_1 (u_2 v_1 - u_1 v_2)/k\right]}{u_1 v_1 + k/2W_1 + u_2 v_2 W_2/W_1 - 2u_1 v_2 W_2 (u_2 v_1 - u_1 v_2)/k} \tag{43}$$

$$G_{12}^d(\omega) \approx \left[ 1 - \left( \frac{\alpha_1}{R_1^3} + \frac{\alpha_2}{R_2^3} \right) \frac{1 - \dfrac{\alpha_1 \alpha_2}{\alpha_1 R_2^3 + \alpha_2 R_1^3} \left( 1 + \dfrac{R_1^3}{R_2^3} \right)}{1 - \dfrac{\alpha_1 \alpha_2}{R_2^6}} \right] \quad (44)$$

### 5. Calculation details

To calculate $dF_{nl}(q,\omega)/d\Omega$ we use the numeric procedures described at length in [7]. We perform calculations in the frame of RPAE approximations. As concrete objects, we choose outer $np^6$ and subvalent $ns^2$ subshells of Ar and Xe and $4d^{10}$ subshell of Xe. We perform computations using equations (5-9, 11, 13, 15, 22-25), for $q = 0.1$, and $1.0$ and transferred energies up to 15 Ry. To illustrate the situation with onion-type endohedrals, we consider $3p^6$ subshell of Ar in Ar@$C_{60}$@$C_{240}$.

Most prominent are the non-dipole corrections at so-called magic angle $\theta_m$, for which the following relation holds $P_2(\cos\theta_m) = 0$. This is why we present differential GOS $dF_{nl}(q,\omega)/d\Omega$ at the magic angle $\theta_m \cong 54.7^0$ and $q = 0.1; 1.0$. We give results also for dipole and non-dipole angular anisotropy parameters. Fig.1-11 collects all data obtained. We employ the following parameters in our calculations, all in atomic units: $R_C = R_{f1} = 6.72$, $R_{f2} = 13.5$, $W_0 = W_1 = 0.4425$, $W_2 = 0.5293$

The lower value of $q$ corresponds to the photoionization limit, since $qR \ll 1$ and in the considered frequency range $\omega/c < 0.05 < q_{\min} = 0.1$. The last inequality shows that we consider also non-dipole corrections to the GOSes that are much bigger than the non-dipole corrections to photoionization.

### 6. Calculation results

We employ the following system of notations in presenting the results of calculations. If one calculates endohedral GOS using the amplitude (8) with all $G_L(\omega,q) = 1$, it is denoted as FRPAE. If the total amplitude is (8), we name corresponding results as GFRPAE. In calculations, having in mind to present semi-qualitative results, we put all $G_L(\omega,q)$ except $G_1(\omega,q)$, equal to one. As to the dipole polarization factor, we neglect its $q$ dependence, so that $G_1(\omega,q) = G(\omega)$.

The results demonstrate that the GOSes and angular anisotropy parameters are complex and informative functions with a number of prominent variations. All calculated characteristics demonstrate strong influence of the electron correlations for $p$-, $s$-, and $d$-electrons. They depend strongly upon the outgoing electron energy and the linear momentum $q$ transferred to an atom in the fast electron inelastic scattering. Electron correlations strongly affect them.

In Figs. 1 and 2 we present differential GOS $dF_{nl}(q,\omega)/d\Omega$ given by (22) and (18) at the magic angle $\theta_m \cong 54.7^0$ [$P_2(\cos\theta_m) = 0$] for outer $np$- and $ns$ - subshells of Ar@$C_{60}$ and Xe@$C_{60}$ at q=0.1 and 1.0 in RPAE, FRPAE and GFRPAE. At q=0.1 the GOS are similar to the photoionization cross-section. The effect of polarization strongly increases the $p$-subshell cross-section value. With growth of q, the maximum decreases in magnitude and shifts to higher $\omega$. Note the double maximum structure that appears due to polarization factor in (8). In Xe at 7 Ry a powerful structure appears due to atomic giant resonance action. For *s*- subshell, the differential GOS values are by an order of magnitude smaller than for s-subshell.

Fig. 3 and 4 depict non-dipole angular anisotropy parameters of knocked-out electrons $\gamma_{ns}^{(in)}(q,\omega)$ and $\eta_{ns}^{(in)}(q,\omega)$ given by (15) [see also (21)] at q=0.1 and 1.0 for subvalent subshell of Ar@$C_{60}$ and Xe@$C_{60}$ in RPAE, FRPAE and GFRPAE. The effect of reflection and polarization corrections is much smaller than in Fig. 1 and 2, but quite visible. Qualitatively, the parameters at q=0.1 are similar to photoionization case. With increase of q the parameters become smaller and variations become broader and less pronounced.

Fig. 5 present dipole angular anisotropy parameters $\beta_{np}^{(in)}(q,\omega)$ of knocked-out electrons given by (22) [see also (26)] at q=0.1 and 1.0 for 3p and 5p subshell of Ar@$C_{60}$ and Xe@$C_{60}$, respectively, in RPAE, FRPAE and GFRPAE. The role of reflection becomes noticeable only at big q. The parameter $\beta_{np}^{(in)}(\omega)$ increases with q growth. The maximum in it goes to higher energies.

Fig. 6 and 7 demonstrate non-dipole angular anisotropy parameters of knocked-out electrons $\gamma_{np}^{(in)}(q,\omega)$ and $\eta_{np}^{(in)}(q,\omega)$ given by (22) [see also (27, 28)] at q=0.1 and 1.0 for 3p subshells of Ar@$C_{60}$ and Xe@$C_{60}$ in RPAE and FRPAE. The structure of $\gamma_{np}^{(in)}(\omega)$ and $\eta_{np}^{(in)}(\omega)$ is very complex and affected by electron correlations, reflection and polarization [see (8)].

Fig. 8 depicts dipole angular anisotropy parameters $\beta_{4d}^{(in)}(q,\omega)$ of knocked-out electrons given by (24) [see also (29)] at q=0.1 and 1.0 for 4d subshell of Xe@$C_{60}$ in RPAE, FRPAE and GFRPAE. The role of reflection becomes noticeable only at big q. The parameter $\beta_{4d}^{(in)}(q,\omega)$ increases with q growth. The role of reflection is noticeable. As to reflection corrections, at these $\omega$ they are negligible.

Fig. 9 shows non-dipole angular anisotropy parameters of knocked-out electrons $\gamma_{4d}^{(in)}(q,\omega)$ and $\eta_{4d}^{(in)}(q,\omega)$ given by (22) [see also (27, 28)], at q=0.1 and 1.0 for 4d subshells of Xe@$C_{60}$ in RPAE and FRPAE. The structure of $\gamma_{4d}^{(in)}(q,\omega)$ and $\eta_{4d}^{(in)}(q,\omega)$ is very complex and affected by electron correlations, reflection and polarization [see (8)].

Fig. 10 gives non-dipole angular anisotropy parameters of knocked-out electrons $\zeta_{nl}^{(in)}(q,\omega)$ determined by (22), at q=0.1 and 1.0 for 3p Ar@$C_{60}$, 5p and 4d Xe@$C_{60}$ subshells in RPAE and FRPAE. The structure in $\zeta_{nl}^{(in)}(q,\omega)$ is quite complex and affected by electron correlations and reflection [see (8)]. This parameter strongly increases with q growth.

Fig. 11 demonstrates differential GOS $dF_{nl}(q,\omega)/d\Omega$ given by (22) and (18) at the magic angle $\theta_m \cong 54.7^0$ [$P_2(\cos\theta_m)=0$] for 3p- subshell of Ar@$C_{60}$@$C_{240}$ at q=0.1 and 1.0 in RPAE, FRPAE and GFRPAE. Here particularly strong is the influence of polarization of both fullerenes shells, $C_{60}$ and $C_{240}$ that increases with q growth.

### 7. Concluding remarks

The main achievement of this paper is demonstration of strong influence of fullerenes shell upon GOS, namely important role played by reflection of knocked-out electrons and by modification of the interaction between incoming electron and endohedral due to polarization of the fullerenes shell. These effects lead to prominent oscillations and new big maxima.

It appeared that GOS and angular anisotropy parameters for endohedral atoms strongly depend upon q and $\omega$. The difference between the angular anisotropy parameters for fast electron scattering and respective photoionization values is big. Already from photoionization studies, we know that atomic electron correlations strongly affect these parameters. Here we saw that fast electron scattering upon endohedrals is very sensitive to transferred momentum. So, their investigation with provide data on these dependences, fullerenes shell effects and their interplay with electron correlations in the central atom.

The biggest unexpected feature of the angular anisotropy for inelastic scattering is that even in the limit of small q they do not coincide with respective photoionization values, and they are not connected to them by simple relation similar to that between photoionization cross-section and GOSes. As we discussed, this is a result of different operators for photoionization and fast electron scattering.

We would be glad to expect that this paper would lead directly to efforts of experimentalists and respective data will appear soon. However, we understand the enormous difficulties of such experiments that require rather dense target of endohedrals in gaseous phase. We understand that such studies require coincidence experiments, in which simultaneously the transferred by fast electron energy and momentum are fixed, as well as the momentum of the secondary electron. However, we do believe that this paper will stimulate interest in such research. The hope is that the interest to the possibility to obtain important results will help to find a way to overcome experimental difficulties. The information that could come from studies of angular distribution of secondary electrons knocked-out endohedrals is of great interest and value. Thus, the suggested here experimental studies are desirable.

Particular and first attention deserves the small q limit. Already the dipole angular anisotropy parameters are different by sign and value. The non-dipole parameters in their turn deviate even qualitatively from their respective photoionization values. It is amazing that in the non-relativistic domain of energies inessential at first glance difference between a virtual and a real photon leads to so prominent consequences.

**Acknowledgements**

The authors are grateful for the financial assistance via the Israeli-Russian grant RFBR-MSTI 11-02-92484

**References**

1. Dolmatov, V. K. in *Theory of Confined Quantum Systems*: *Part Two*, J. R. Sabin and E. Brändas, Ed., Advances in Quantum Chemistry (Academic Press, New York, 2009, Vol. **58**, 13.
2. Mitsuke, K.; Mori, T; Kou, J. *The Journal of Chemical Physics* **2005**, 122, 064304 1-5.
3. Connerade, J.-P.; Dolmatov, V. K.; Manson, S. T. *J. Phys. B: At. Mol. Opt. Phys*. **2000** 33, 2279-2285
4. Amusia, M. Ya.; Baltenkov, A. S. *Phys. Rev*. 2006, A 73, 062723-1-6.
5. Amusia, M. Ya. *Nanotubes, and Carbon Nanostructures* 2010, 18, 353 – 368.
6. Amusia, M. Ya., *Atomic Photoeffect*, N. Y.-London, Plenum Press, 1990.
7. Amusia M. Ya.; Chernysheva L.V., *Computation of Atomic Processes*, Institute of Physics Publishing, Bristol and Philadelphia, 1997
8. Amusia M. Ya.; Baltenkov, A. S.; Chernysheva, L.V.; Felfli, Z.; Msezane, A. Z. *Phys. Rev*. A 2001, **63**, 052506-1-7.
9. El-Sherbini T. M., Van der Wiel M. J., *Physica,* 1972
10. Amusia, M. Ya.; Chernysheva, L. V.; Liverts, E. Z. on line 2011, *Phys. Rev.* A, 2012, submitted, 2012

11. Inokuti, M. *Rev. Mod. Phys.* 1971, 43, 297.
12. Amusia, M. Ya.; Chernysheva L. V.; Felfli Z.; Msezane A. Z., *Phys. Rev.* A 2003, 67, 022703-1-8.
13. Amusia, M. Ya., *Radiation Physics and Chemistry* 2004, 70, 237-251.
14. http://wolfram.com
15. Baltenkov, A. S. *J. Phys. B: At. Mol. Opt. Phys.* 1999, 38, L169.
16. Amusia, M. Ya.; Baltenkov, A. S.; Chernysheva, L. V.; Felfli, Z.; Msezane, A. Z., *J. Phys. B: At. Mol. Opt. Phys.* 2005, 38, L169-73.
17. Kilcoyne, A. L. D.; Aguilar, A.; Möller, A.; Schippers, S.; Cisneros, C.; Alna'Washi, G.; Aryal, N. B.; Baral, K.K.; Esteves, D. A.; Thomas, C. M.; Phaneuf, R. A. *Phys. Rev. Lett.* 2010, 105, 213001-1-4.
18. Amusia, M. Ya.; Chernysheva, L. V.; Liverts, E. Z. *JETP Letters* 2011, 94, 431-436.
19. Berkowitz, J.; *J. Chem. Phys.* 1999, 111, 1446-53.
20. Forró, L.; Mihály, L. *Rep. Prog. Phys.*, 2001, **64**, 649
21. Amusia, M. Ya.; Chernysheva, L. V.; Liverts, E. Z. *Phys. Rev. A* 2009, 80.032503-1-12.

**Figure captions**

Fig.1. Differential GOS $dF_{np}(q,\omega)/d\Omega$ at the magic angle $P_2(\cos\theta_m)=0$ of 3p- and 5p- subshells in Ar@$C_{60}$ and Xe@$C_{60}$, respectively, at q=0.1 and q=1.0 in RPAE, FRPAE, and GFRPAE.

Fig.2. Differential GOS $dF_{ns}(q,\omega)/d\Omega$ at the magic angle $P_2(\cos\theta_m)=0$ of 3s- and 5p- subshells in Ar@$C_{60}$ and Xe@$C_{60}$, respectively, at q=0.1 and q=1.0 in RPAE, FRPAE, and GFRPAE.

Fig.3. Angular anisotropy non-dipole parameters $\gamma_{3s}^{(in)}(q,\omega)$ and $\eta_{3s}^{(in)}(q,\omega)$ of knocked-out electrons in fast projectile-atom collision at q=0.1 and q=1.0 for 3s- subshell of Ar@$C_{60}$ in RPAE, FRPAE, and GFRPAE.

Fig.4. Angular anisotropy non-dipole parameters $\gamma_{5s}^{(in)}(q,\omega)$ and $\eta_{5s}^{(in)}(q,\omega)$ of knocked-out electrons in fast projectile-atom collision at q=0.1 and q=1.0 for 3s- subshell of Xe@$C_{60}$ in RPAE, FRPAE, and GFRPAE.

Fig. 5. Dipole angular anisotropy parameters $\beta_{np}^{(in)}(q,\omega)$ of knocked-out electrons given by (22) [see also (26)] at q=0.1 and 1.0 for 3p and 5p subshell of Ar@$C_{60}$ and Xe@$C_{60}$, respectively, in RPAE, FRPAE and GFRPAE.

Fig.6 Non-dipole angular anisotropy parameters of knocked-out electrons $\gamma_{np}^{(in)}(q,\omega)$ and $\eta_{3p}^{(in)}(q,\omega)$ given by (22) [see also (27, 28)] at q=0.1 and 1.0 for 3p subshells of Ar@$C_{60}$ in RPAE, FRPAE and GFRPAE.

Fig.7. Non-dipole angular anisotropy parameters of knocked-out electrons $\gamma_{5p}^{(in)}(q,\omega)$ and $\eta_{5p}^{(in)}(q,\omega)$ given by (22) [see also (27, 28)] at q=0.1 and 1.0 for 3p subshells of Xe@$C_{60}$ in RPAE, FRPAE and GFRPAE.

Fig. 8. Angular anisotropy dipole parameters $\beta_{4d}^{(in)}(q,\omega)$ of knocked-out electrons in fast projectile-atom collision with 4d- subshell in Xe@$C_{60}$, at q=0.1 and q=1.0 in RPAE, and FRPAE.

Fig.9. Angular anisotropy non-dipole parameters $\gamma_{4d}^{(in)}(q,\omega)$ and $\eta_{4d}^{(in)}(q,\omega)$ of knocked-out electrons in fast projectile-atom collision at q=0.1 and q=1.0 for 4d- subshell of Xe@$C_{60}$ in RPAE, FRPAE, and GFRPAE.

Fig.10. Angular anisotropy non-dipole parameter $\zeta_{nl}^{(in)}(q,\omega)$ of knocked-out electrons at q=0.1 and q=1.0 for 3p Ar@$C_{60}$, 5p- and 4d Xe@$C_{60}$ subshells, respectively, in RPAE, FRPAE, and GFRPAE.

Fig. 11. Differential GOS $dF_{nl}(q,\omega)/d\Omega$ given by (22) and (18) at the magic angle $\theta_m \cong 54.7^0$ [$P_2(\cos\theta_m)=0$] for 3p- subshell of Ar@$C_{60}$@$C_{240}$ at q=0.1 and 1.0 in RPAE, FRPAE and GFRPAE..

**Figures**

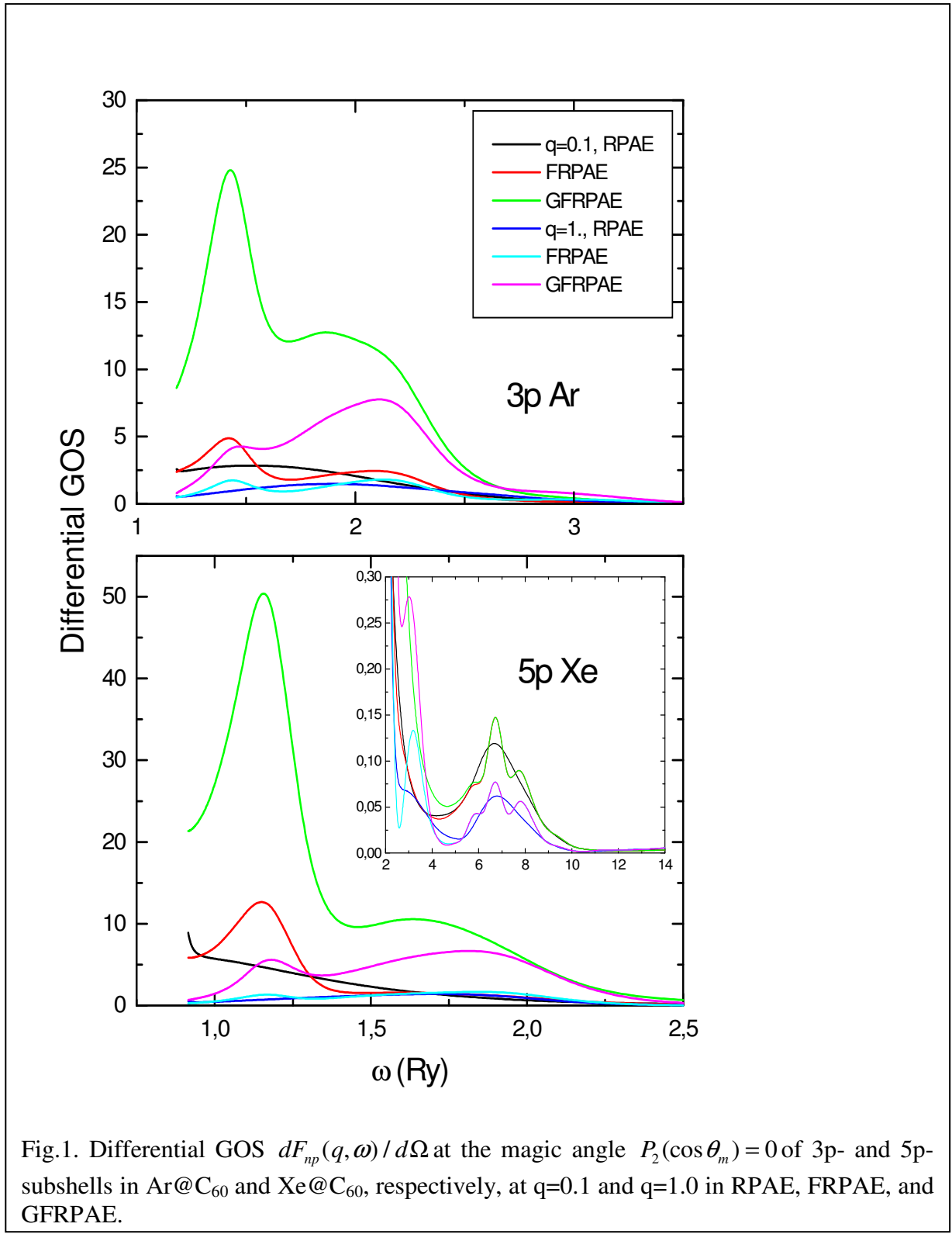

Fig.1. Differential GOS $dF_{np}(q,\omega)/d\Omega$ at the magic angle $P_2(\cos\theta_m)=0$ of 3p- and 5p-subshells in Ar@$C_{60}$ and Xe@$C_{60}$, respectively, at q=0.1 and q=1.0 in RPAE, FRPAE, and GFRPAE.

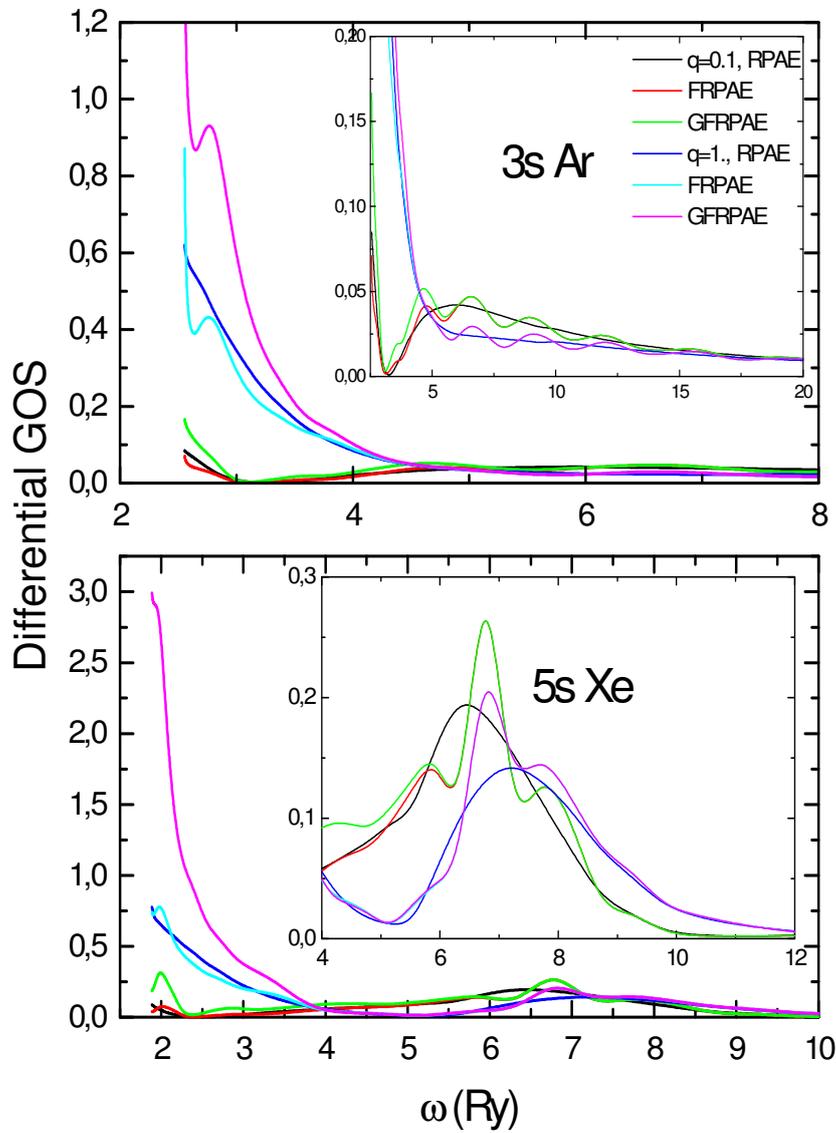

Fig.2. Differential GOS $dF_{ns}(q,\omega)/d\Omega$ at the magic angle $P_2(\cos\theta_m)=0$ of 3s- and 5p- subshells in Ar@$C_{60}$ and Xe@$C_{60}$, respectively, at q=0.1 and q=1.0 in RPAE, FRPAE, and GFRPAE.

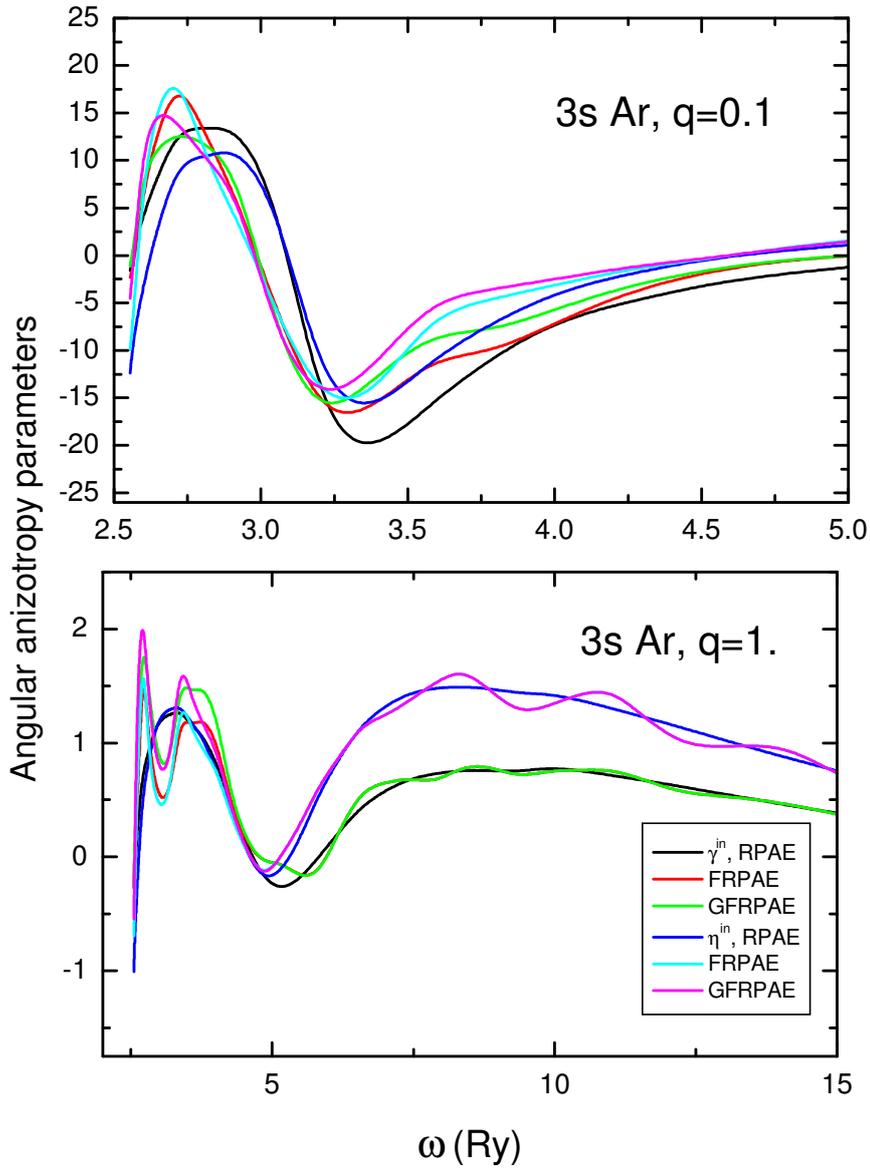

Fig.3. Angular anisotropy non-dipole parameters $\gamma_{3s}^{(in)}(q,\omega)$ and $\eta_{3s}^{(in)}(q,\omega)$ of knocked-out electrons in fast projectile-atom collision at q=0.1 and q=1.0 for 3s- subshell of Ar@$C_{60}$ in RPAE, FRPAE, and GFRPAE.

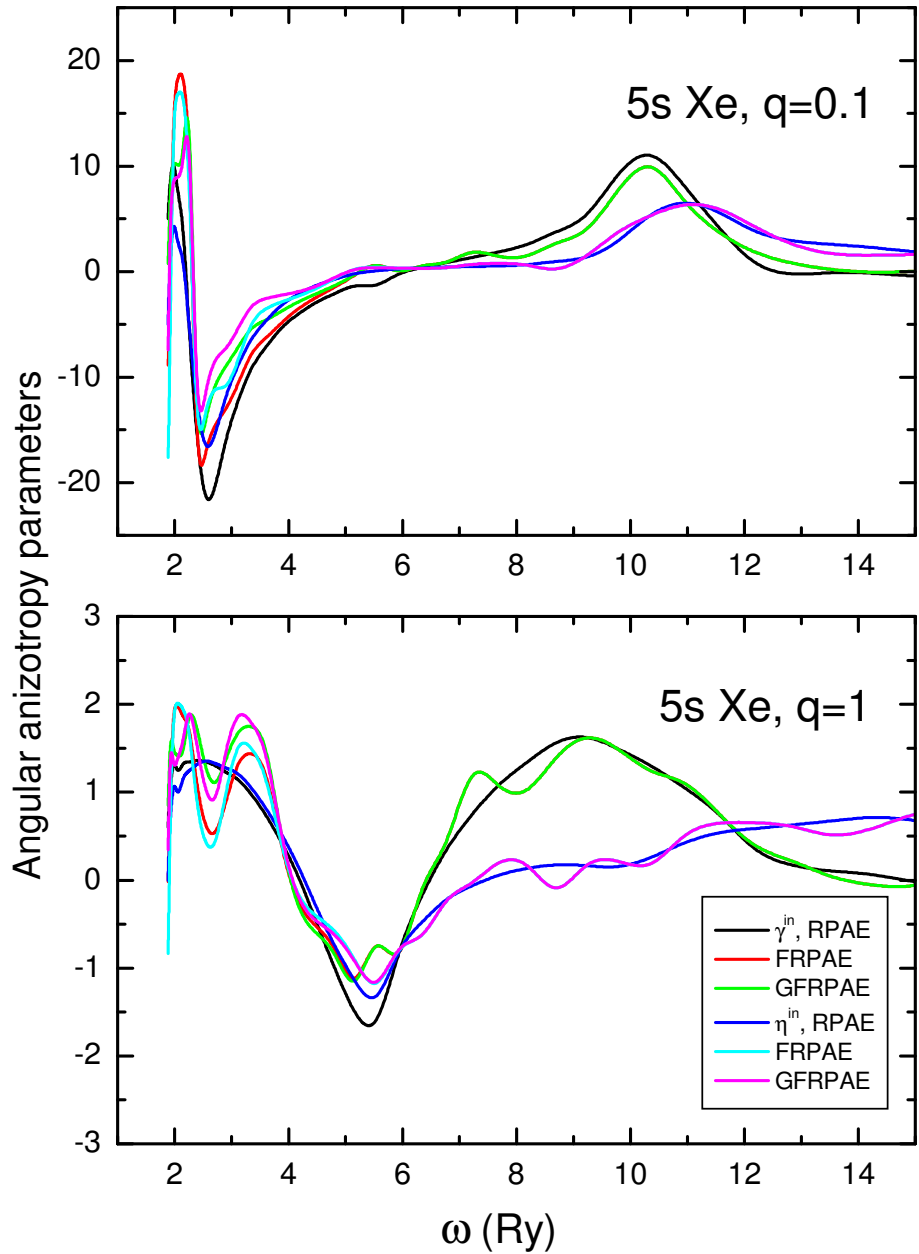

Fig.4. Angular anisotropy non-dipole parameters $\gamma_{5s}^{(in)}(q,\omega)$ and $\eta_{5s}^{(in)}(q,\omega)$ of knocked-out electrons in fast projectile-atom collision at q=0.1 and q=1.0 for 3s- subshell of Xe@C$_{60}$ in RPAE, FRPAE, and GFRPAE.

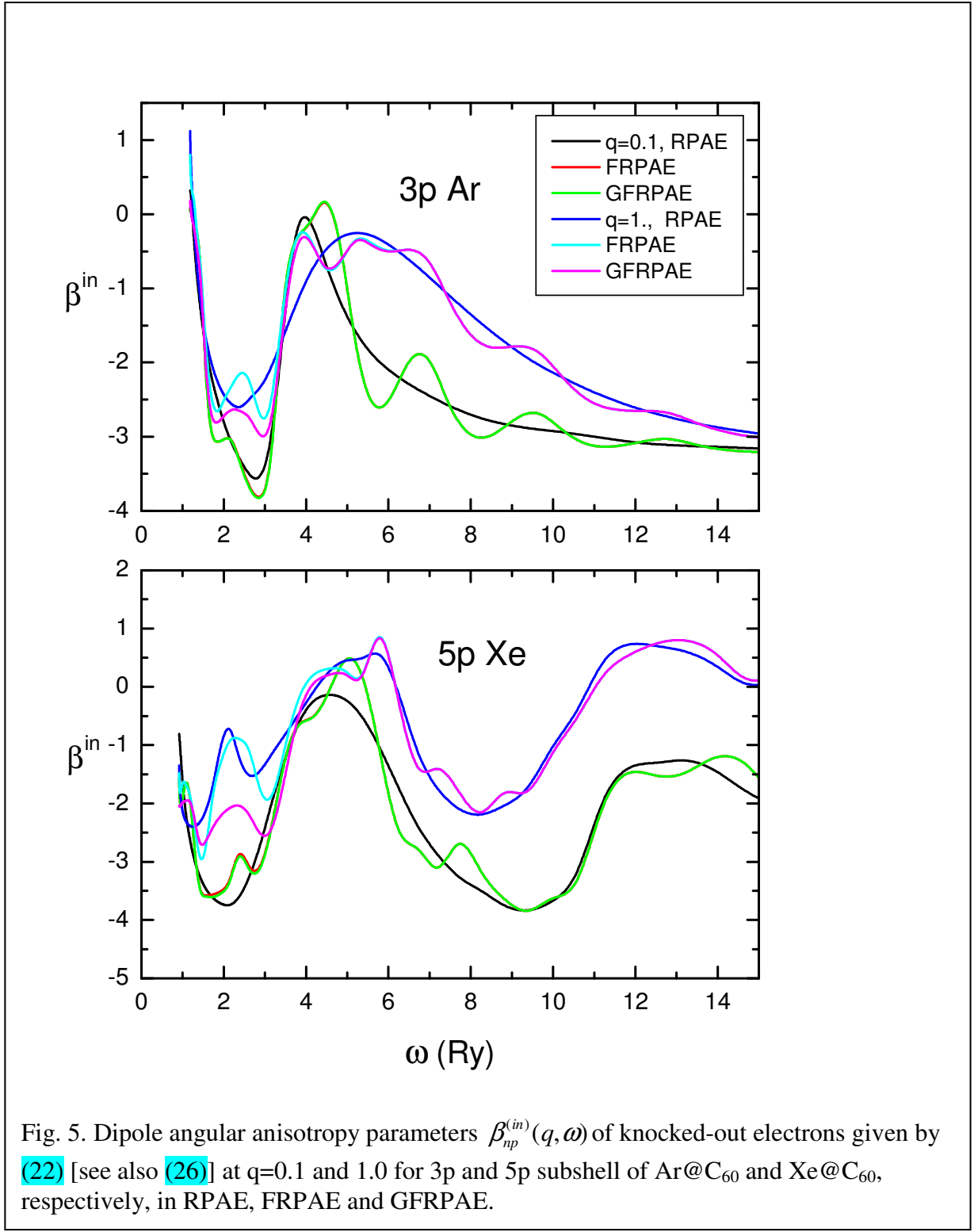

Fig. 5. Dipole angular anisotropy parameters $\beta_{np}^{(in)}(q,\omega)$ of knocked-out electrons given by (22) [see also (26)] at q=0.1 and 1.0 for 3p and 5p subshell of Ar@$C_{60}$ and Xe@$C_{60}$, respectively, in RPAE, FRPAE and GFRPAE.

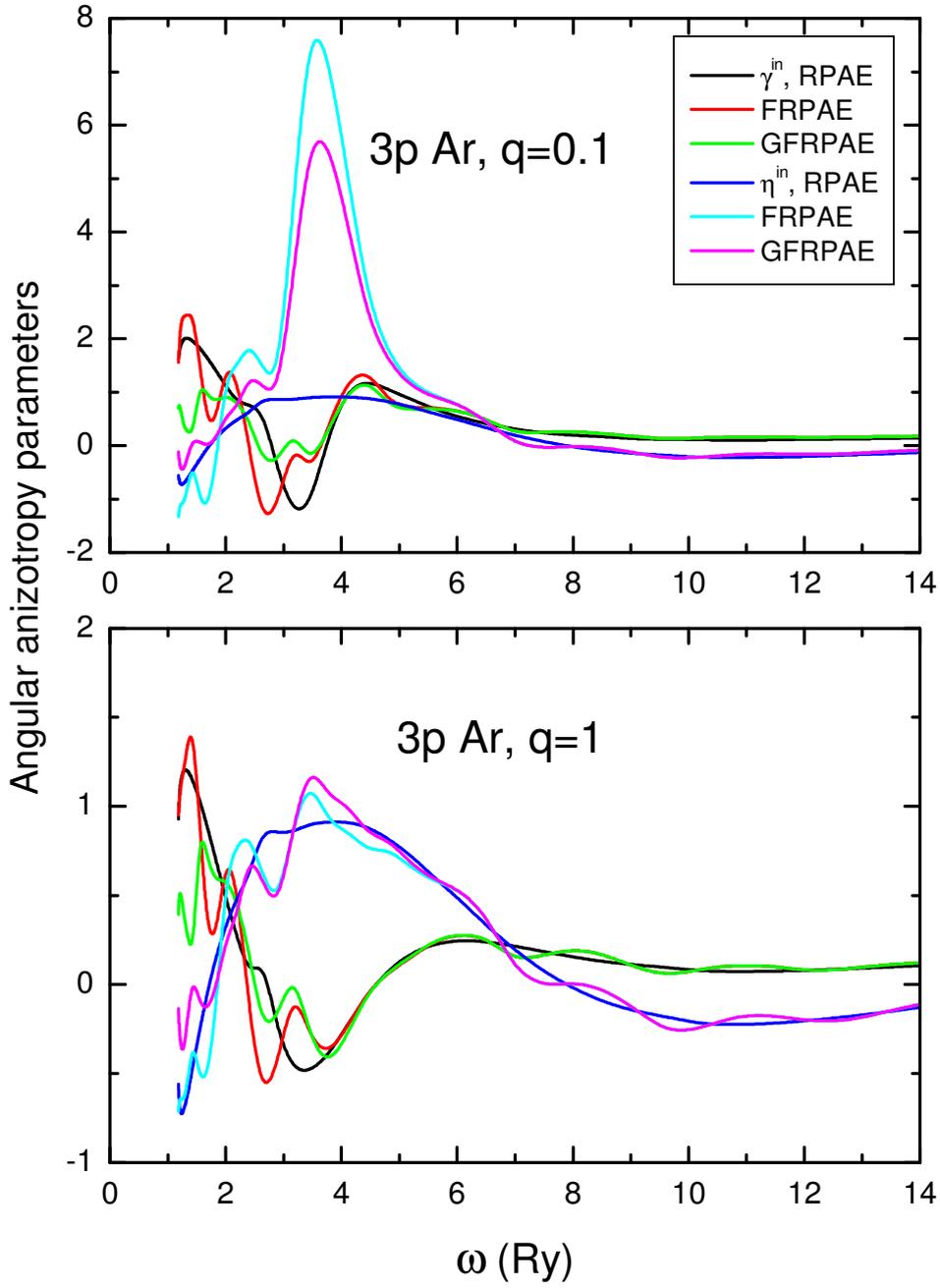

Fig.6 Non-dipole angular anisotropy parameters of knocked-out electrons $\gamma_{np}^{(in)}(q,\omega)$ and $\eta_{3p}^{(in)}(q,\omega)$ given by (22) [see also (27, 28)] at q=0.1 and 1.0 for 3p subshells of Ar@$C_{60}$ in RPAE, FRPAE and GFRPAE.

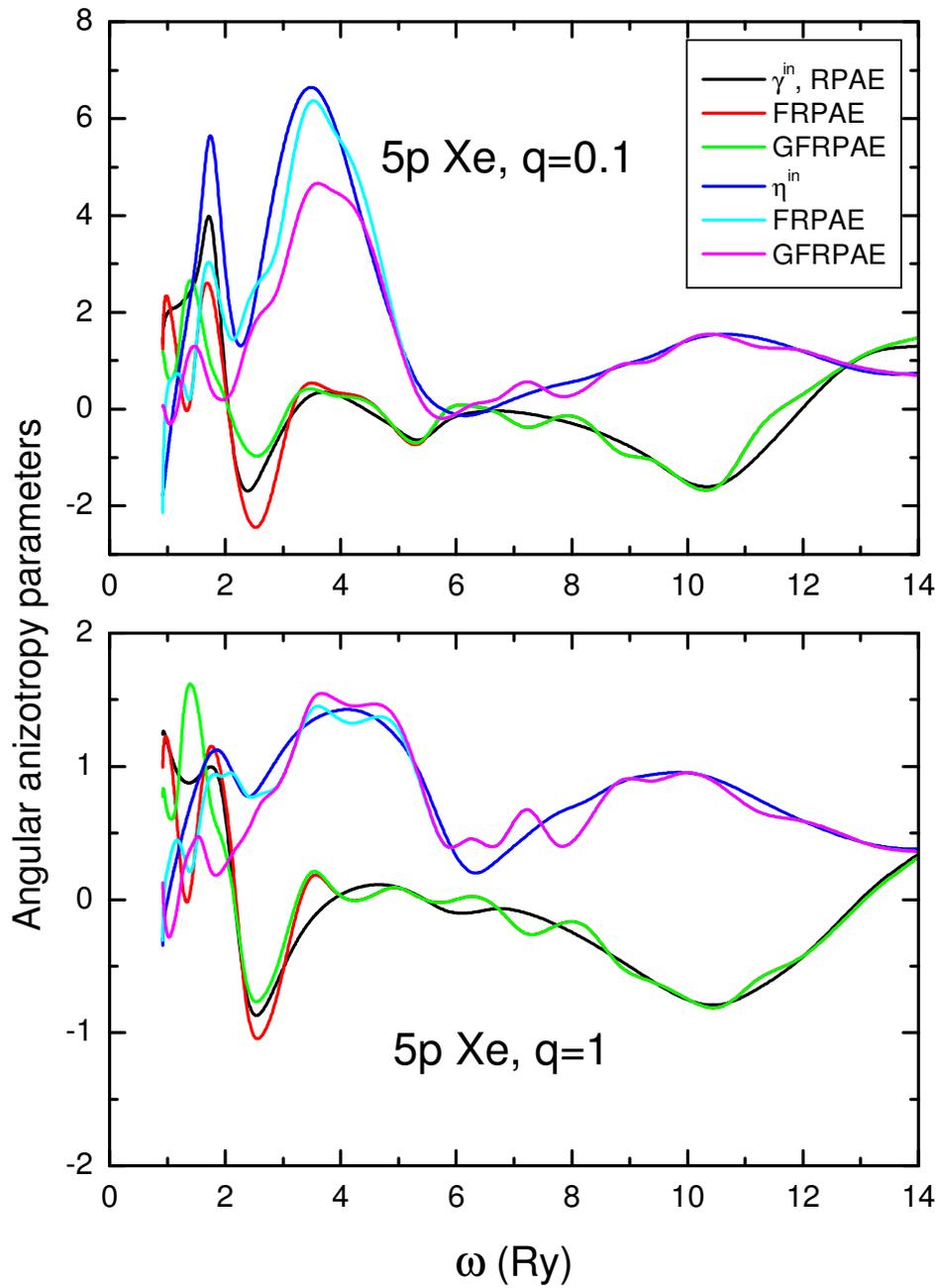

Fig.7. Non-dipole angular anisotropy parameters of knocked-out electrons $\gamma_{5p}^{(in)}(q,\omega)$ and $\eta_{5p}^{(in)}(q,\omega)$ given by (22) [see also (27, 28)] at q=0.1 and 1.0 for 3p subshells of Xe@C$_{60}$ in RPAE, FRPAE and GFRPAE.

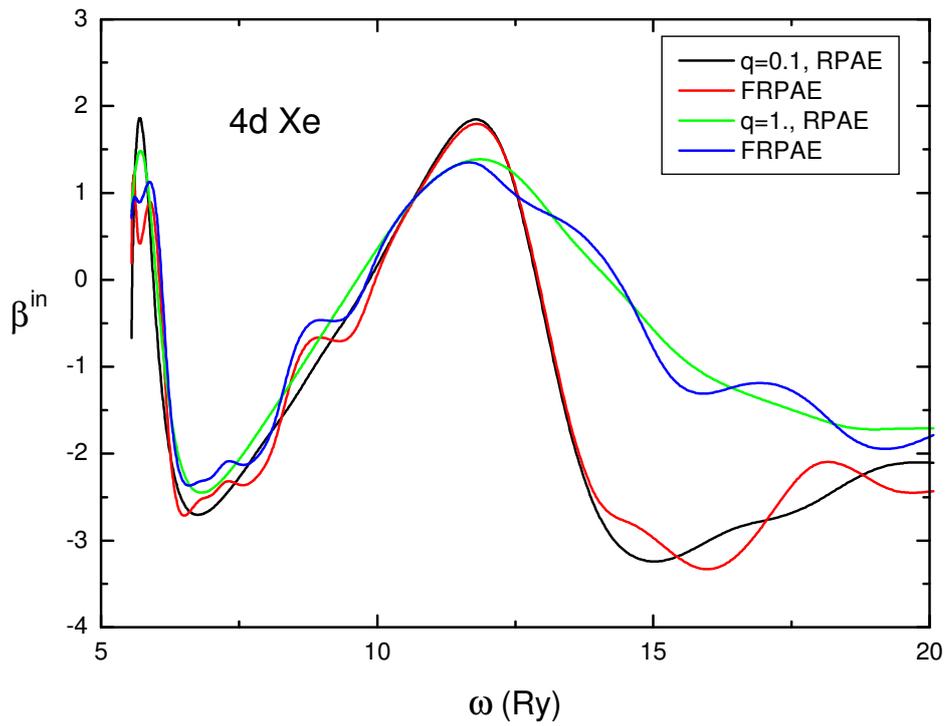

Fig. 8. Angular anisotropy dipole parameters $\beta_{4d}^{(in)}(q,\omega)$ of knocked-out electrons in fast projectile-atom collision with 4d- subshell in Xe@$C_{60}$, at q=0.1 and q=1.0 in RPAE, and FRPAE.

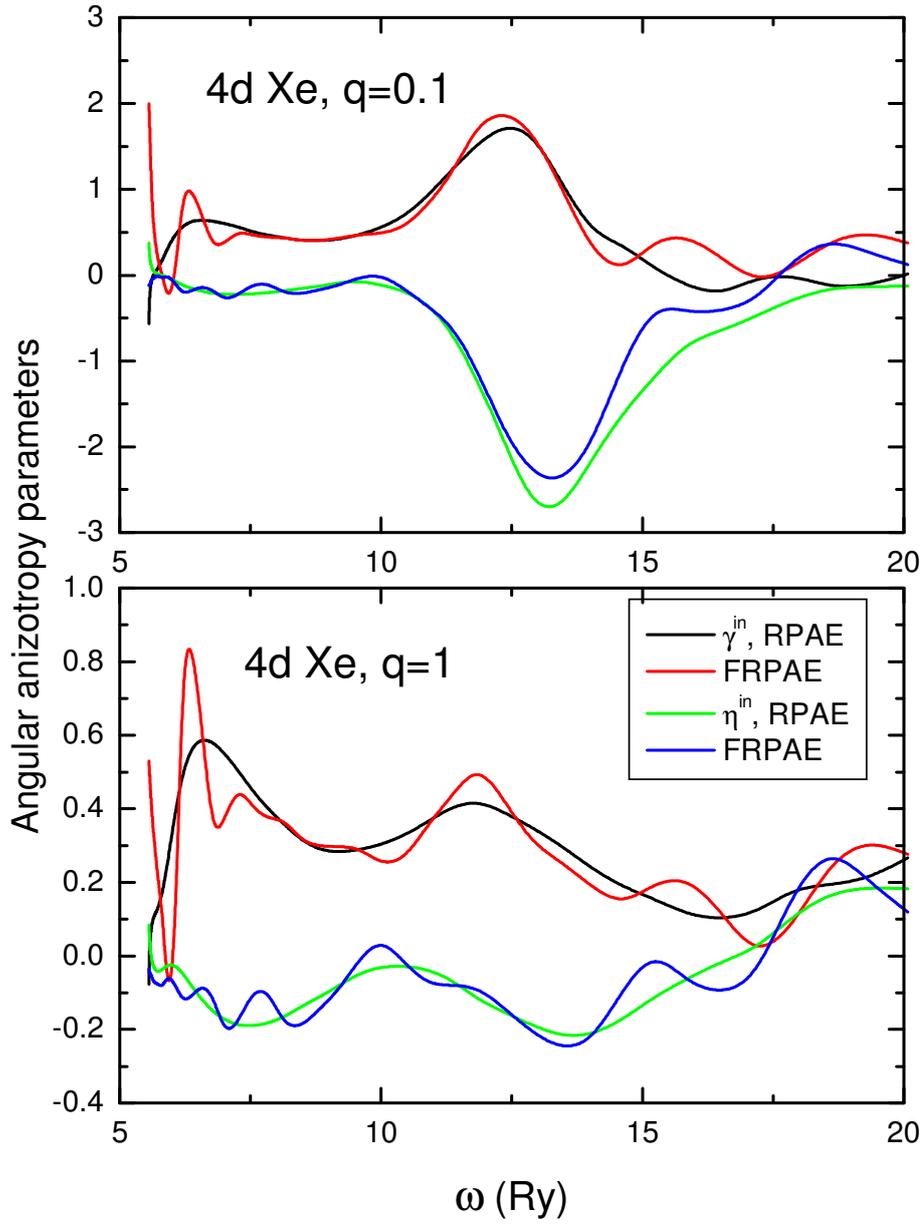

Fig.9. Angular anisotropy non-dipole parameters $\gamma_{4d}^{(in)}(q,\omega)$ and $\eta_{4d}^{(in)}(q,\omega)$ of knocked-out electrons in fast projectile-atom collision at q=0.1 and q=1.0 for 4d- subshell of Xe@$C_{60}$ in RPAE, FRPAE, and GFRPAE.

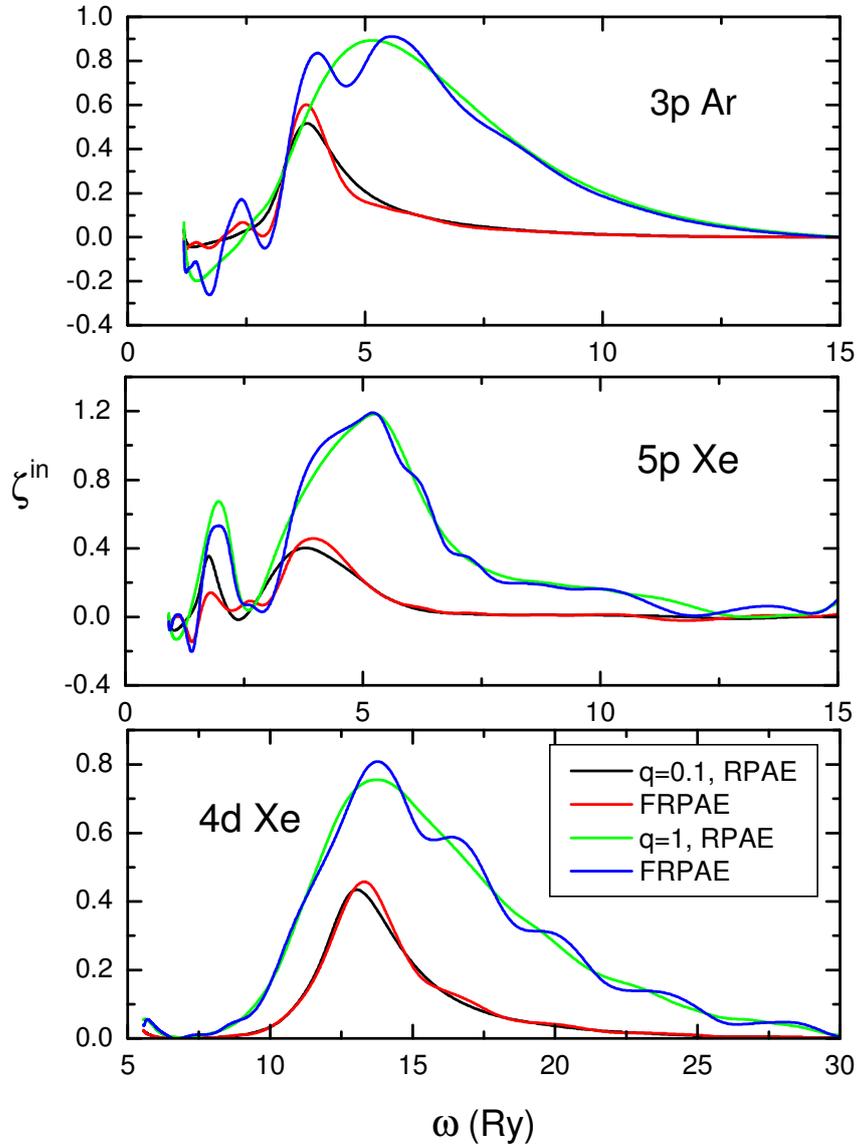

Fig 10. Angular anisotropy non-dipole parameter $\zeta_{nl}^{(in)}(q,\omega)$ of knocked-out electrons at q=0.1 and q=1.0 for 3p Ar@$C_{60}$, 5p- and 4d Xe@$C_{60}$ subshells, respectively, in RPAE, FRPAE, and GFRPAE.

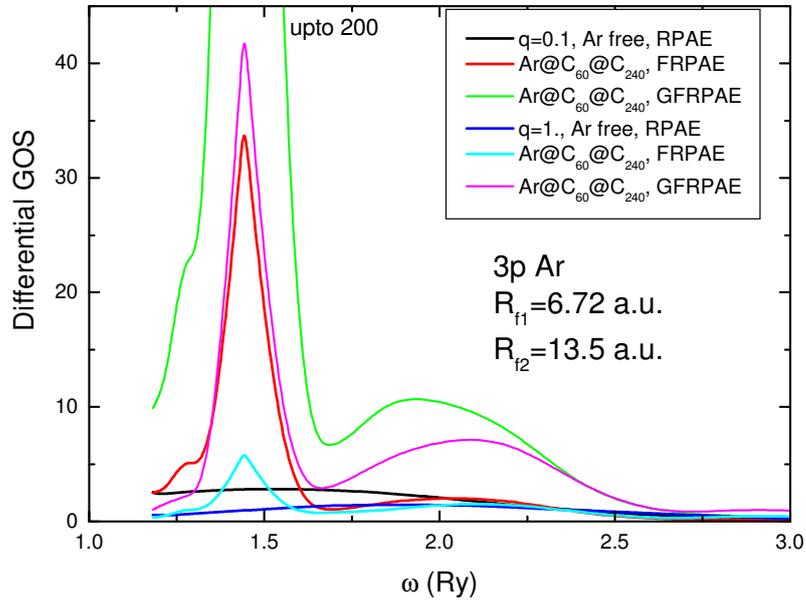

Fig. 11. Differential GOS $dF_{nl}(q,\omega)/d\Omega$ given by (22) and (18) at the magic angle $\theta_m \cong 54.7^0$ [$P_2(\cos\theta_m) = 0$] for $3p$- subshell of Ar@C$_{60}$@C$_{240}$ at q=0.1 and 1.0 in RPAE, FRPAE and GFRPAE..